\documentclass[twoside,11pt]{article}

\usepackage{blindtext}

\usepackage{jmlr2e}

\usepackage{hyperref}
\usepackage{amsmath,amsfonts,amssymb}
\usepackage{url}
\usepackage{booktabs} 
\usepackage{xfrac}
\usepackage{graphicx}
\usepackage{wrapfig}
\usepackage{adjustbox}
\usepackage{capt-of}
\usepackage{xfrac}
\usepackage{lipsum}
\usepackage{enumitem}
\usepackage[caption=false]{subfig}
\usepackage[ruled,vlined]{algorithm2e}
\usepackage{algorithmic}
\newcommand\norm[1]{\left\lVert#1\right\rVert}
\usepackage{float}

\usepackage{lastpage}

\begin{document}

\title{Off-Policy Action Anticipation in Multi-Agent Reinforcement Learning}

\author{\name Ariyan Bighashdel \email a.bighashdel@tue.nl \\
        \name Daan de Geus \email d.c.d.geus@tue.nl \\
        \name Pavol Jancura \email p.jancura@tue.nl \\
        \name Gijs Dubbelman \email g.dubbelman@tue.nl \\
       \addr Department of Electrical Engineering\\
       Eindhoven University of Technology\\
       Eindhoven, 5612 AZ, The Netherlands}


\maketitle

\begin{abstract}

Learning anticipation in Multi-Agent Reinforcement Learning (MARL) is a reasoning paradigm where agents anticipate the learning steps of other agents to improve cooperation among themselves. As MARL uses gradient-based optimization, learning anticipation requires using Higher-Order Gradients (HOG), with so-called HOG methods. Existing HOG methods are based on \textit{policy parameter anticipation}, i.e., agents anticipate the changes in policy parameters of other agents. Currently, however, these existing HOG methods have only been applied to differentiable games or games with small state spaces. In this work, we demonstrate that in the case of non-differentiable games with large state spaces, existing HOG methods do not perform well and are inefficient due to their inherent limitations related to policy parameter anticipation and multiple sampling stages. To overcome these problems, we propose Off-Policy Action Anticipation (OffPA2), a novel framework that approaches learning anticipation through action anticipation, i.e., agents anticipate the changes in actions of other agents, via off-policy sampling. We theoretically analyze our proposed OffPA2 and employ it to develop multiple HOG methods that are applicable to non-differentiable games with large state spaces. We conduct a large set of experiments and illustrate that our proposed HOG methods outperform the existing ones regarding efficiency and performance. 
\end{abstract}


\section{Introduction}
\label{sec:Introduction}
\begin{figure}[t]
  \centering
  \begin{minipage}[c]{1\textwidth}
    \centering
    \subfloat{\label{fig:gamesetting}\includegraphics[width=1\textwidth]{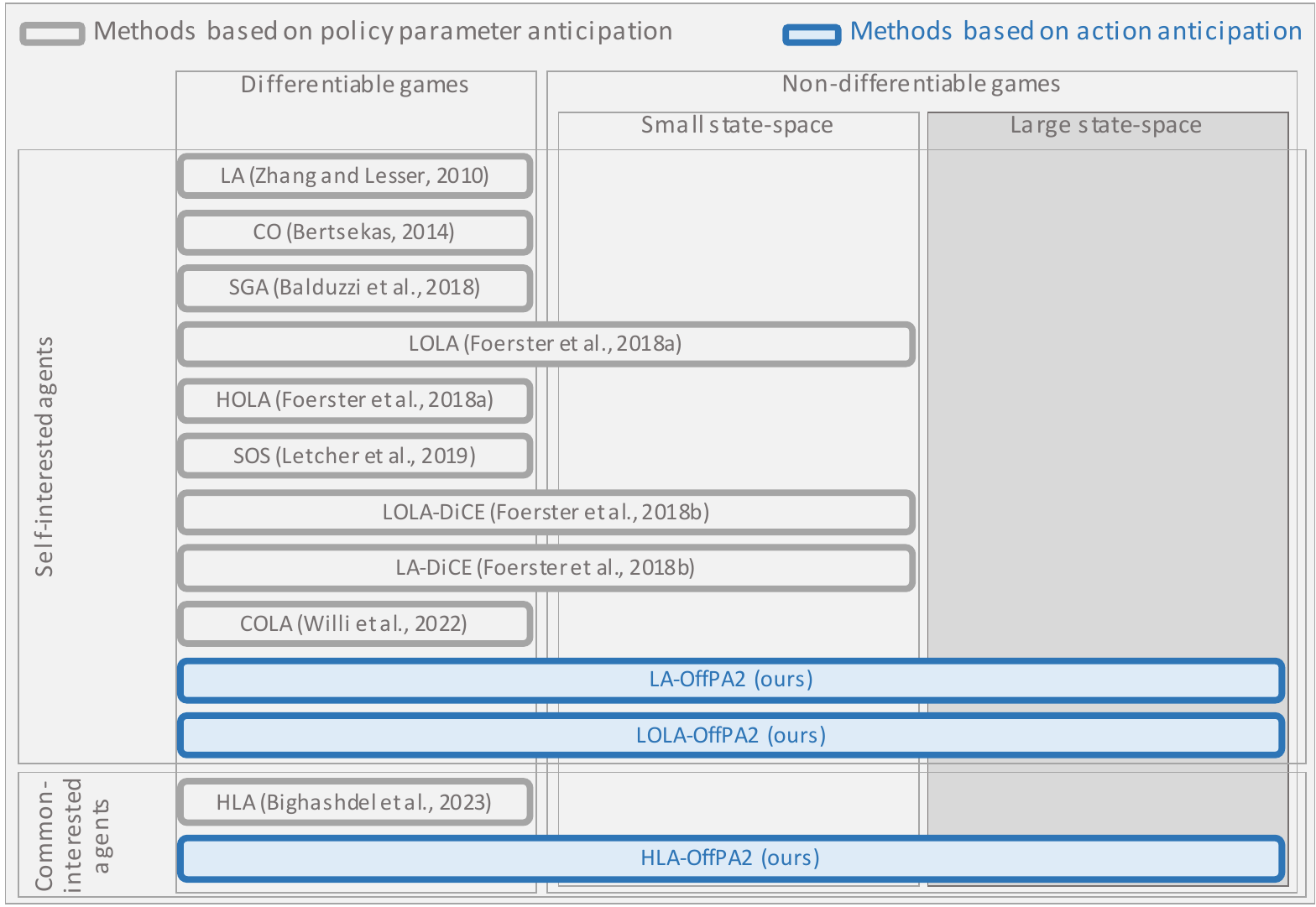}}%
    \captionof{figure}{\label{fig:taxonomy} Overview of HOG methods and their applicability in various game settings. Existing HOG methods (gray rectangles) are based on the policy parameter approach. These methods have been only applied to differentiable games or non-differentiable games with small state spaces. Our proposed HOG methods (blue rectangles) are developed in our novel framework of Off-Policy Action Anticipation (OffPA2) and can be applied to non-differentiable games with large state spaces.}
  \end{minipage}
\end{figure}

In multi-agent systems, the paradigm of \textit{agents' reasoning about other agents} has been explored and researched extensively~\citep{goodie2012levels,liu2021reasoning}. Recently, this paradigm is also being studied in the subfield of Multi-Agent Reinforcement Learning (MARL) \citep{wen2019probabilistic,wen2020modelling,konan2022iterated}. Generally speaking, MARL deals with several agents simultaneously learning and interacting in an environment. In the context of MARL, one reasoning strategy is anticipating the learning steps of other agents \citep{zhang2010multi}, i.e., learning anticipation. As MARL uses gradient-based optimization, learning anticipation naturally leads to the usage of Higher-Order Gradients (HOG), with so-called HOG methods \citep{letcher2018stable}. The significance of learning anticipation in HOG methods has been frequently shown in the literature. For instance, Look-Ahead (LA) \citep{zhang2010multi,letcher2018stable} uses learning anticipation to guarantee convergence in cyclic games such as matching pennies, Learning with Opponent-Learning Awareness (LOLA) \citep{foerster2018learning} employs learning anticipation to ensure cooperation in general-sum games such as Iterated Prisoner's Dilemma (IPD), and Hierarchical Learning anticipation (HLA) \citep{Bighashdel2023} utilizes learning anticipation to improve coordination among common-interested agents in fully-cooperative games. In this study, we explore the limitations of current HOG methods and propose novel solutions so that learning anticipation can be applied to a broader range of MARL problems. In Figure \ref{fig:taxonomy}, we provide an overview of the applicability of both existing and our proposed HOG methods.



Learning anticipation in the current HOG methods is developed based on the \textit{policy parameter anticipation} approach, i.e., agents anticipate the changes in policy parameters of other agents \citep{zhang2010multi,foerster2018learning,foerster2018dice} (see Figure \ref{fig:taxonomy}). In this approach, first of all, agents should either have access to other agents' exact parameters or infer other agents' parameters from state-action trajectories \citep{foerster2018learning}. In many game settings, these parameters are obscured. This is problematic because when the size of the state space increases, the dimensionality of the parameter spaces increases as well, making the parameter inference problem computationally expensive. Furthermore, anticipating the changes in high-dimensional policy parameters is inefficient, whether the parameters are inferred or exact. Finally, policy parameter anticipation requires higher-order gradients with respect to policy parameters which is shown to be challenging in MARL~\citep{foerster2018dice,lu2022model}. Current HOG methods mainly assume that the games are differentiable, i.e., agents have access to gradients
and Hessians~\citep{willi2022cola,letcher2018stable} (Figure \ref{fig:gamesetting}). When the games are non-differentiable, existing HOG methods employ the Stochastic Policy Gradient (SPG) theorem \citep{sutton2018reinforcement} with on-policy sampling to compute the gradients with respect to the policy parameters~\citep{foerster2018learning,foerster2018dice}. However, estimating higher-order gradients in SPG requires either analytical approximations -- since the learning step for one agent in the standard SPG theorem is independent of other agents' parameters -- or multi-stage sampling, which is inefficient and comes typically with high variance, making learning unstable~\citep{foerster2018learning,foerster2018dice}. In this work, we aim to propose novel HOG methods that overcome the aforementioned limitations of existing HOG methods, making learning anticipation applicable to non-differentiable games with large state spaces.

To accomplish our goal, we propose Off-Policy Action Anticipation (OffPA2), a novel framework that approaches learning anticipation through action anticipation (see Figure \ref{fig:taxonomy}). Specifically, the agents in OffPA2 anticipate the changes in actions of other agents during learning. Unlike policy parameter anticipation, action anticipation is performed in the action space whose dimensionality is generally lower than the policy parameter space in MARL games with large state spaces~\citep{lowe2017multi,Peng2021}. Furthermore, we employ the Deterministic Policy Gradient (DPG) theorem with off-policy sampling to estimate differentiable objective functions. Consequently, high-order gradients can be efficiently computed, while still following the standard Centralized Training and Decentralized Execution (CTDE) setting where agents can observe the other agents' actions during training~\citep{lowe2017multi}. We theoretically analyze our OffPA2 in terms of performance and time complexity. The proposed OffPA2 framework allows us to develop HOG methods that, unlike existing HOG methods, are applicable to non-differentiable games with large state spaces. To show this, we apply the principles of LA, LOLA, and HLA to our OffPA2 framework and develop the LA-OffPA2, LOLA-OffPA2, and HLA-OffPA2 methods, respectively. We compare our methods with existing HOG methods in well-controlled studies. By doing so, we demonstrate that the overall performance and efficiency of our proposed methods do not dot decrease with increasing the state-space size, unlike for existing HOG methods, where they get drastically worse. Finally, we compare our methods with the standard, DPG-based MARL algorithms and highlight the importance of learning anticipation in MARL. Below, we summarize our contributions.
\begin{itemize}
    \item We propose OffPA2, a novel framework that approaches learning anticipation through action anticipation, which makes HOG methods applicable to non-differentiable games with large state spaces. We provide theoretical analyses of the influence of our proposed action anticipation approach on performance and time complexity.
    \item  Within our OffPA2 framework, we develop three novel methods, i.e., LA-OffPA2, LOLA-OffPA2, and HLA-OffPA2. We show that our methods outperform the existing HOG methods and state-of-the-art DPG-based approaches.
\end{itemize}

\section{Related work}
\label{sec:Related works}

In many real-world MARL tasks, communication constraints during execution require the use of decentralized policies. In these cases, one reasoning tool is Agents Modeling Agents (AMA) \citep{albrecht2018autonomous}, where agents explicitly model other agents to predict their behaviors. Although AMA traditionally assumes na\"ive opponents with no reasoning abilities \citep{he2016opponent,hong2018deep}, recent studies have extended AMA to further consider multiple levels of reasoning where each agent considers the reasoning process of other agents to make better decisions \citep{wen2019probabilistic,wen2020modelling}. For instance, \cite{wen2019probabilistic} proposed the probabilistic recursive reasoning (PR2) update rule for MARL agents to recursively reason about other agents' beliefs.  However, in these approaches, agents do not take into account the learning steps of other agents, which has shown to be important in games where interaction among self-interested agents otherwise leads to worst-case outcomes \citep{foerster2018learning}. In Section \ref{sec:Experiments}, we conduct several experiments and compare our proposed methods with these approaches.

HOG methods, on the other hand, are a range of methods that use higher-order gradients to consider the anticipated learning steps of other agents. These include: 1) LOLA and Higher-order LOLA (HOLA), proposed by \cite{foerster2018learning} to improve cooperation in Iterated Prisoner's Dilemma (IPD), 2) Look-Ahead (LA), proposed by \cite{zhang2010multi} to guarantee convergence in cyclic games, 3) Stable Opponent Shaping (SOS), developed by \cite{letcher2018stable} as an interpolation between LOLA and LA to inherit the benefits of both, 4) Consistent LOLA (COLA), proposed by \cite{willi2022cola} to improve consistency in opponent shaping, 5) Hierarchical Learning Anticipation (HLA), proposed by \cite{Bighashdel2023} to improve coordination among fully-cooperative agents, 6) Consensus Optimization (CO), proposed by \cite{bertsekas2014constrained} to improve training stability and convergence in zero-sum games, and 7) Symplectic Gradient Adjustment (SGA), proposed by \cite{balduzzi2018mechanics} to improve parameter flexibility of CO.


Despite their novel ideas, most existing HOG methods have been only applied to differentiable games, where the agents have access to the exact gradients or Hessians, and only LOLA has been evaluated on non-differentiable games. Specifically, \cite{foerster2018learning} employed the SPG framework to estimate the gradients in LOLA. As the standard SPG is independent of other agents' parameters, the authors relied on Taylor expansions of the expected return combined with analytical derivations of the second-order gradients. \cite{foerster2018dice} indicated that this approach is not stable in learning. To solve the problem, \cite{foerster2018dice} proposed an infinitely differentiable Monte Carlo estimator, referred to as DiCE, to correctly optimize the stochastic objectives with any order of gradients. Similarly to meta-learning, the agents in the DiCE framework reason about and predict the learning steps of the opponents using inner learning loops and update their parameters in outer learning loops. However, each learning loop for each agent requires a sampling stage which is very inefficient for high-order reasoning and games with large state spaces, i.e., beyond matrix games. In Section \ref{sec:Experiments}, we conduct a set of experiments to closely compare our proposed OffPA2 framework with DiCE.

\section{Problem formulation and background}
\label{sec:Background}
We formulate the MARL setup as a Markov Game (MG) \citep{littman1994markov}. An MG is a tuple $(\mathcal{N},\mathcal{S},\{\mathcal{A}_i\}_{i \in \mathcal{N}},\{\mathcal{R}_i\}_{i \in \mathcal{N}},\mathcal{T},\rho,\gamma)$, where $\mathcal{N}$ is the set of agents ($|\mathcal{N}|=n$), $\mathcal{S}$ is the set of states, and $\mathcal{A}_i$ is the set of possible actions for agent $i \in \mathcal{N}$. Agent $i$ chooses its action $a_i \in \mathcal{A}_i$ through the stochastic policy network $\pi_{\theta_i}:\mathcal{S} \times \mathcal{A}_i \rightarrow [0,1]$ parameterized by $\theta_i$ conditioning on the given state $s \in \mathcal{S}$. Given the actions of all agents, each agent $i$ obtains a reward $r_i$ according to its reward function $\mathcal{R}_i:\mathcal{S}\times \mathcal{A}_1 \times ... \times \mathcal{A}_{n} \rightarrow \mathbb{R}$. Given an initial state, the next state is produced according to the state transition function $\mathcal{T}:\mathcal{S} \times \mathcal{A}_1 \times ... \times \mathcal{A}_{n} \times \mathcal{S} \rightarrow [0,1]$. We denote an episode of horizon $T$ as $\tau=(\{s^0,a^0_1,...,a^0_{n},r^0_1,...,r^0_{n}\},...,\{s^T,a^T_1,...,a^T_{n},r^T_1,...,r^T_{n}\})$, and the discounted return for each agent $i$ at time step $t \leq T$ is defined by $G^t_i(\tau)=\sum_{l=t}^T \gamma^{l-t} r_i$ where $\gamma$ is a predefined discount factor. The expected return given the agents’ policy parameters approximates the state value function for each agent  $V_i(s,\theta_1,...,\theta_{n})=\mathbb{E}[G_i^t(\tau|s^t=s)]$. The goal for each agent $i$ is to find the policy parameters, $\theta_i$, that maximize the expected return given the distribution of the initial state $\rho(s)$, denoted by the performance objective $J_i = \mathbb{E}_{\rho(s)} V_i(s,\theta_1,...,\theta_{n})$. 

\textbf{Na\"ive gradient ascend}. In the na\"ive update rule, agents do not perform learning anticipation to update their policy parameters. More specifically, each na\"ive agent $i$ maximizes its performance objective by updating its policy parameters in the direction of the objective's gradient
\begin{equation}
\begin{split}
    \nabla_{\theta_i} J_i = \mathbb{E}_{\rho(s)}\nabla_{\theta_i} V_i(s,\theta_1,...,\theta_{n}).
\end{split}
\end{equation}

\textbf{Learning With Opponent-Learning Awareness (LOLA)}. Unlike na\"ive agents, LOLA agents modify their learning objectives by differentiating through the anticipated learning steps of the opponents \citep{foerster2018learning}. Given $n=2$ for simplicity, a first-order LOLA agent (agent One) assumes a na\"ive opponent and uses policy parameter anticipation to optimize $V_1^{\text{LOLA}}(s,\theta_1,\theta_2+\Delta\theta_2)$ where $\Delta\theta_2 = \mathbb{E}_{\rho(s)}  \eta\nabla_{\theta_2}V_2(s,\theta_1,\theta_2)$ and $\eta\in \mathbb{R}^+$ is the prediction length. Using first-order Taylor expansion and by differentiating with respect to $\theta_1$, the gradient adjustment for the first LOLA agent~\citep{foerster2018learning}~is given by
\begin{equation}
\label{eq:lola}
\begin{split}
    \nabla_{\theta_1}V_1^{\text{LOLA}}(s,\theta_1,\theta_2+\Delta\theta_2) \approx \nabla_{\theta_1}V_1 + (\nabla_{\theta_2\theta_1}V_1)^\intercal \Delta\theta_2 + \underbrace{ (\nabla_{\theta_1} \Delta\theta_2)^\intercal \nabla_{\theta_2}V_1}_{\text{shaping}},
\end{split}
\end{equation}
where $V_1 = V_1(s,\theta_1,\theta_2)$. The rightmost term in the LOLA update allows for active shaping of the opponent's learning. This term has been proven effective in enforcing cooperation in various games, including IPD \citep{foerster2018learning,foerster2018dice}. The LOLA update can be further extended to non-na\"ive opponents, resulting in HOLA agents \citep{foerster2018learning,willi2022cola}.

\textbf{Look Ahead (LA).} LA agents assume that the opponents' learning steps cannot be influenced, i.e., cannot be shaped \citep{zhang2010multi,letcher2018stable}. In other words, agent One assumes that the prediction step, $\Delta{\theta}_2$, is independent of the current optimization, i.e., $\nabla_{\theta_1} \Delta{\theta}_2 = 0$. Therefore, the shaping term disappears, and the gradient adjustment for the first LA agent will be
\begin{equation}
\label{eq:la}
\begin{split}
    \nabla_{\theta_1}V_1^{\text{LA}}(s,\theta_1,{\theta}_2+\perp\Delta{\theta}_2) \approx \nabla_{\theta_1}V_1 + (\nabla_{{\theta}_2\theta_1}V_1)^\intercal \Delta{\theta}_2,
\end{split}
\end{equation}
where $\perp$ prevents gradient flowing from $\Delta{\theta}_2$ upon differentiation. 

\textbf{Hierarchical Learning Anticipation (HLA).} Unlike LOLA and LA, HLA is proposed to improve coordination in fully cooperative games with common interested agents \citep{Bighashdel2023}, i.e., $\mathcal{R}_i=\mathcal{R}_j=\mathcal{R} \;\forall i,j \in \mathcal{N}$ and, consequently, $V_i=V_j=V \;\forall i,j \in \mathcal{N}$. HLA randomly assigns the agents into hierarchy levels to specify their reasoning orders. In each hierarchy level, the assigned agent is a \textit{leader} of the lower hierarchy levels and a \textit{follower} of the higher ones, with two reasoning rules: 1) a leader knows the reasoning levels of the followers and is one level higher, and 2) a follower cannot shape the leaders and only follows their shaping plans. Concretely, if $n=2$, and we assume that agent Two is the leader (HLA-L) and agent One is the follower (HLA-F), the gradient adjustment for the leader is:
\begin{equation}
\label{eq:hr_leader}
\begin{split}
    \nabla_{\theta_2}V^{\text{HLA-L}}(s,\theta_1+\Delta\theta_1,\theta_2) \approx \nabla_{\theta_2}V + (\nabla_{\theta_1\theta_2}V)^\intercal \Delta\theta_1 + (\nabla_{\theta_2} \Delta\theta_1)^\intercal \nabla_{\theta_1}V,
\end{split}
\end{equation}
where $V = V(s,\theta_1,\theta_2)$ is the common value function, and $\Delta\theta_1 =  \eta\nabla_{\theta_1}V$. The plan of the leader is to change its parameters $\bar{\theta}_2=\theta_2 + \eta\nabla_{\theta_2}V^{\text{HLA-L}}(s,\theta_1+\Delta\theta_1,\theta_2)$ in such a way that an optimal increase in the common value is achieved after its new parameters are taken into account by the follower. Therefore, the follower must follow the plan and adjust its parameters through 
\begin{equation}
\label{eq:hr_follower}
\begin{split}
    \nabla_{\theta_1}V^{\text{HLA-F}}(s,\theta_1,\bar{\theta}_2) \approx \nabla_{\theta_1}V + (\nabla_{\theta_2\theta_1}V)^\intercal \eta\nabla_{\theta_2}V^{\text{HLA-L}}(s,\theta_1+\Delta\theta_1,\theta_2).
\end{split}
\end{equation}

\section{Approach}
\label{sec:HOG high-dimensional}

In this section, we propose OffPA2, a framework designed to enable the application of HOG methods to non-differentiable games with large state spaces. To solve the problems regarding policy parameter anticipation, we propose the novel approach of action anticipation, where agents anticipate the changes in actions of other agents during learning. Furthermore, we employ the DPG theorem with off-policy sampling to estimate differentiable objective functions. Consequently, high-order gradients can be efficiently computed. Our proposed OffPA2 complies with the standard Centralized Training and Decentralized Execution (CTDE) setting in DPG, where the agents during training have access to the actions of other agents~\citep{lowe2017multi}.


\subsection{OffPA2: Off-policy action anticipation}
We define a deterministic policy $\mu_{\theta_i}: \mathcal{S} \rightarrow \mathcal{A}_i$, parameterized by $\theta_i$ for each agent $i\in \mathcal{N}$. Let $Q_i(s,a_1,...,a_{n})=\mathbb{E}[G_i^t(\tau|s^t=s,a_i^t=a_i \forall i \in \mathcal{N})]$ denote the state-action value function, then we have  $V_i(s,\theta_1,...,\theta_{n})= Q_i(s,\mu_{\theta_1}(s),..., \mu_{\theta_n}(s))$. Furthermore, we define a stochastic behavior policy for each agent $i$ as $\pi^b_{\theta_i} = \mu_{\theta_i}+SG$, where $SG$ is a standard Gaussian distribution. Given the behavior policies, the policy parameters can be learned off-policy, from trajectories generated by the behavior policies, i.e., $\rho^b(s,\{a_i\}_{i\in\mathcal{N}},\{r_i\}_{i\in\mathcal{N}},s')$ where $s$ and $s'$ are consecutive states. Using the deterministic policy gradient theorem \citep{silver2014deterministic,lowe2017multi}, we can obtain the gradient of the performance objective for each na\"ive agent $i$ as $\nabla_{\theta_i}J_i  = \mathbb{E}_{\rho^\beta(s)} \nabla_{\theta_i} Q_i(s,\mu_{\theta_1}(s) ,...,\mu_{\theta_n}(s) )$.

Without the loss of generality, we consider two agents, i.e., $n=2$, and we assume that agent One wants to anticipate the learning step of agent Two, who is a na\"ive learner. At each state $s \sim \rho^b(s)$, agent One anticipates the changes in the policy parameters of agent Two as $\Delta \theta_2 (s)= \eta \nabla_{\theta_2}Q_2(s,\mu_{\theta_1}(s),\mu_{\theta_2}(s))$, i.e., policy parameter anticipation. Therefore, agent One updates its policy parameters in the direction of: 
\begin{equation}
\label{eq:gradient_agent2}
\begin{split}
    \nabla_{\theta_1}J_1 = \mathbb{E}_{\rho^\beta(s)} \nabla_{\theta_1}Q_1(s,\mu_{\theta_1},\mu_{\theta_2+\Delta \theta_2(s)}(s)),
\end{split}
\end{equation}

\begin{theorem}[action anticipation]
\label{prop:action anticipation}
Using first-order Taylor expansion, the gradient of the performance objective for agent One, Eq. (\ref{eq:gradient_agent2}), can be approximated as:
\begin{equation}
\label{eq:gradient_agent2_action}
\begin{split}
    \nabla_{\theta_1}J_1 \approx \mathbb{E}_{\rho^\beta(s)} \nabla_{\theta_1}\mu_{\theta_1}(s)\nabla_{a_1}Q_1(s,a_1,a_2+\Delta a_2)|_{a_1=\mu_{\theta_1}(s),a_2=\mu_{\theta_2}(s)},
\end{split}
\end{equation}
where 
\begin{equation}
\begin{split}
    \Delta a_2=\hat{\eta}_{\text{1st}}\nabla_{a_2}Q_2(s,a_1,a_2),
\end{split}
\end{equation}
is the anticipated change of action, where $\hat{\eta}_{\text{1st}}=\eta \norm{\nabla_{\theta_2}\mu_{\theta_2(s)}}^2\in \mathbb{R}^+$ is the projected prediction length.
\end{theorem}
{\bf Proof}. See Appendix \ref{apsec:Proof of Proposition action}.

Theorem \ref{prop:action anticipation} indicates that agent One can anticipate the learning step of agent Two in the action space rather than the policy parameter space. This way of reasoning has two benefits. First, in the MARL games with large state spaces, the dimensionality of action space is significantly lower than that of the policy parameter space ~\citep{lowe2017multi,Peng2021}. The justification is that large state spaces require more complex policy networks with more parameters to properly represent all possible states. Second, action anticipation, unlike policy parameter anticipation, complies with the standard centralized training and decentralized execution (CTDE) setting in DPG. In the standard CTDE setting, the agents during training have access to the centralized state-action value functions to train the decentralized policies. Consequently, the agents are informed of other agents' actions and can perform action anticipation during training. This is while in policy parameter anticipation, the agents need to additionally access the policy parameters of other agents.

\subsubsection{Influence of action anticipation on performance}
Our proposed action anticipation approach employs the first-order Taylor approximation to map the anticipated learning from the policy parameter space to the action space. In other words:
\begin{equation}
\begin{split}
     \mu_{{\theta}_2+\Delta{\theta}_2(s)}(s) \approx a_2 + \Delta a_2,
\end{split}
\end{equation}
where $a_2=\mu_{{\theta}_2}(s)$ and $\Delta a_2 = \hat{\eta}_{\text{1st}}\nabla_{a_2}Q_2(s,a_1,a_2)$. In the theorem below, we show how this approximation influences the principles of HOG methods, which have been theoretically and experimentally researched throughout the literature \citep{zhang2010multi,foerster2018learning,letcher2018stable,willi2022cola}. 

\begin{theorem}
\label{th:projection_performance}
For a sufficiently small $\hat{\eta}_{\text{1st}}$, there exists an $\eta'\in \mathbb{R}^+$ such that 
\begin{equation}
\begin{split}
     \mu_{{\theta}_2+\Delta{\theta'}_2(s)}(s) = a_2 +  \Delta a_2,
\end{split}
\end{equation}
where
\begin{equation}
\begin{split}
     &\Delta \theta'_2 (s)= \eta' \nabla_{\theta_2}\mu_{\theta_2}(s) \nabla_{a_2} Q_2(s,a_1,a_2)\\
     &a_2=\mu_{{\theta}_2}(s)\\
     &\Delta a_2 = \hat{\eta}_{\text{1st}}\nabla_{a_2}Q_2(s,a_1,a_2)\;\;\;\;\;\;\;\hat{\eta}_{\text{1st}}=\eta \norm{\nabla_{\theta_2}\mu_{\theta_2(s)}}^2\;\;\;\;\;\;\;\eta \in \mathbb{R}^+
\end{split}
\end{equation}
\end{theorem}
{\bf Proof}. See Appendix \ref{apsec:Proof of Theorem projection}.

Based on Theorem \ref{th:projection_performance}, action anticipation via first-order Taylor expansion and sufficiently small $\hat{\eta}_{\text{1st}}$ only scales the prediction length as both $\eta$ and $\eta'$ are non-negative numbers. The general theoretical analyses on differentiable games reveal that scaling the prediction length influences the HOG methods' convergence behaviors \citep{letcher2018stable,zhang2010multi}. However, in our OffPA2 framework, we directly set the \textit{projected} prediction length, i.e., $\hat{\eta}_{\text{1st}}$, rather than the prediction length, i.e., $\eta$. Consequently, the resulting prediction length in the policy parameter space, i.e., $\eta'$, can correspond to satisfactory convergence behaviors (see Section \ref{sec:Influence of the projection estimation}).

\subsubsection{Computing higher-order gradients}
The state-action value function in Eq (\ref{eq:gradient_agent2_action}) is generally unknown and non-differentiable. Similarly to the DPG-based algorithms~\citep{silver2014deterministic,lowe2017multi}, we substitute a differentiable state-action value function $Q_i(s,a_1,...,a_{n};\omega_i)$, parameterized by $\omega_i$, in place of the true state-action value function, i.e., $Q_i(s,a_1,...,a_{n};\omega_i)\approx Q_i(s,a_1,...,a_{n})$. The parameters of the state-action value function can be obtained by minimizing the Temporal Difference (TD) error, off-policy, from
episodes generated by the behavior policies \citep{lowe2017multi}: 
\begin{equation}
\begin{split}
     \mathcal{L}(\omega_i) = \mathbb{E}_{\rho^b(s,\{a_i\}_{i\in\mathcal{N}},\{r_i\}_{i\in\mathcal{N}},s')}[(Q_i(s,a_1,...,a_{n};\omega_i)-y_i)^2],
\end{split}
\end{equation}
where $y_i$ is the TD target value:
\begin{equation}
\begin{split}
     y_i=r_i+\gamma Q'_i(s,a'_1,...,a'_{n})|_{a'_i=\mu'_i(s')\;\forall i \in \mathcal{N}},
\end{split}
\end{equation}
where $Q'_i$ and $\mu'_i$ are the target state-action value and policy functions, respectively. The differentiability of objective functions in OffPA2 is particularly beneficial for HOG methods as they need to frequently compute the higher-order gradients to anticipate the agents' learning. 

\subsubsection{Influence of action anticipation on time complexity}
Apart from the differentiability of objective functions, the action anticipation approach further reduces the gradient computation complexity as it requires the anticipated changes of actions, i.e., $\Delta a_i$, rather than the anticipated 
changes of policy parameters, i.e., $\Delta\theta_i$. Assume that the policy and state-action value networks are multi-layer perceptrons (as done in most experiments), then:

 \begin{theorem}
\label{th:time}
 Action anticipation, compared to policy parameter anticipation, reduces the time complexity of anticipating the learning step of a na\"ive opponent by $O(L N^2)$, where $L$ is the number of fully connected layers, and $N$ is the number of neurons per layer in the policy and state-action value networks.
\end{theorem}
{\bf Proof}. See Appendix \ref{apsec:Proof of Theorem time}.

\subsection{OffPA2-based HOG methods}
Having the OffPA2 framework, we can now develop HOG methods that are applicable to non-differentiable games with large state spaces. In the following sections, we develop LOLA-OffPA2, LA-OffPA2, and HLA-OffPA2 by applying the LOLA, LA, and HLA principles to our OffPA2 framework, respectively.

\subsubsection{LOLA-OffPA2}
\label{sec:lola-OffPA2}
As described in Section \ref{sec:Background}, LOLA agents predict and shape the learning steps of other agents to improve cooperation in non-team games. Given two agents ($n=2$) for simplicity, the LOLA-OffPA2 agent (agent One) predicts and shapes the action of the opponent (agent Two) that is assumed by agent One to be na\"ive. Using first-order Taylor expansion, the gradient adjustment for the first LOLA-OffPA2 agent is given by
\begin{equation}
\label{eq:LOLA-OffPA2_Taylor}
\begin{split}
    \nabla_{\theta_1}J_1^\text{LOLA-OffPA2} &= \mathbb{E}_{\rho^\beta(s)} \nabla_{\theta_1}\mu_{\theta_1}(s)\nabla_{a_1}Q_1(s,a_1,a_2+\Delta a_2)|_{a_1=\mu_{\theta_1}(s),a_2=\mu_{\theta_2}(s)}\\
    & \approx \mathbb{E}_{\rho^\beta(s)} \nabla_{\theta_1}\mu_{\theta_1}(s)\left( \nabla_{a_1}Q_1 + (\nabla_{a_2 a_1}Q_1)^\intercal \Delta a_2 + \underbrace{(\nabla_{a_1} \Delta a_2)^\intercal \nabla_{a_2}Q_1}_{\text{action shaping}},
    \right)
\end{split}
\end{equation}
where 
\begin{equation}
\label{eq:LOLA-OffPA2_Taylor_add}
\begin{split}
    & \Delta a_2 = \hat{\eta}_{\text{1st}}\nabla_{a_2}Q_2(s,a_1,a_2)|_{a_1=\mu_{\theta_1}(s),a_2=\mu_{\theta_2}(s)}\\
    & Q_1 = Q_1(s,a_1,a_2)|_{a_1=\mu_{\theta_1}(s),a_2=\mu_{\theta_2}(s)}.
\end{split}
\end{equation} 

The rightmost term in the LOLA-OffPA2 update, i.e., Eq. (\ref{eq:LOLA-OffPA2_Taylor}), allows for active \textit{action shaping} of the opponent. 

In practice, we don't need to rely on Taylor expansion for the update rules in LOLA-OffPA2 as we can use an automatic differentiation engine, e.g., PyTorch autograd \citep{paszke2019pytorch}, to directly compute the gradients. Algorithm \ref{alg:LOLA-OffPA2} in Appendix illustrates the LOLA-OffPA2 optimization framework for the case of $n$ agents. At each state $s\sim\rho^b(s)$ the agent $i \in \mathcal{N}$ first anticipates the changes in actions of all agents $j \in \{\mathcal{N}-\{i\}\}$:
\begin{equation}
\label{eq:LOLA-OffPA2_add}
\begin{split}
    \Delta a_j = \hat{\eta}_{\text{1st}}\nabla_{a_2}Q_2(s,a_1,...,a_n)|_{a_i=\mu_{\theta_i}(s)\;\forall i \in \mathcal{N}}.
\end{split}
\end{equation}
Then, agent $i$ updates its parameters $\theta_i$ by the following gradient adjustment:
\begin{equation}
\label{eq:LOLA-OffPA2}
\begin{split}
    \nabla_{\theta_i}J_i^{\text{LOLA-OffPA2}} = \mathbb{E}_{\rho^\beta(s)} \nabla_{\theta_i}\mu_{\theta_i}(s)\nabla_{a_i}Q_i(s,a_1+\Delta a_1,...,a_i,...,a_n+\Delta a_n)|_{a_i=\mu_{\theta_i}(s)\;\forall i \in \mathcal{N}}.
\end{split}
\end{equation}
Equation (\ref{eq:LOLA-OffPA2}) denotes the update rule for \textit{first-order} LOLA-OffPA2 agents that assume na\"ive opponents. However, we can also consider a \textit{second-order} LOLA-OffPA2 agent that differentiates through the learning steps of first-order LOLA-OffPA2 opponents. Likewise, we can extend the update rules to include higher-order reasoning, as in HOLA \citep{foerster2016learning}.

\subsubsection{LA-OffPA2}
\label{sec:la-OffPA2}
Similarly to the LA principles \citep{zhang2010multi,letcher2018stable}, LA-OffPA2 agents cannot shape the opponents' learning steps, i.e., they cannot shape the opponent's actions. Consequently, in the two-agent case, we have $\nabla_{a_1} \Delta a_2=0$. Using first-order Taylor expansion, the gradient adjustment for the first LA-OffPA2 agent~\citep{foerster2018learning}~is given by
\begin{equation}
\label{eq:LA-OffPA2_Taylor}
\begin{split}
    \nabla_{\theta_1}J_1^\text{LA-OffPA2} &= \mathbb{E}_{\rho^\beta(s)} \nabla_{\theta_1}\mu_{\theta_1}(s)\nabla_{a_1}Q_1(s,a_1,a_2+\perp\Delta a_2)|_{a_1=\mu_{\theta_1}(s),a_2=\mu_{\theta_2}(s)}\\
    & \approx \mathbb{E}_{\rho^\beta(s,\hat{a}_2)} \nabla_{\theta_1}\mu_{\theta_1}(s)\left( \nabla_{a_1}Q_1 + (\nabla_{a_2 a_1}Q_1)^\intercal \Delta a_2
    \right),
\end{split}
\end{equation}
where $\perp$ prevents gradient flowing from $\Delta a_2$ upon differentiation, and $\Delta a_2$ and $Q_1$ are defined in Eq. (\ref{eq:LOLA-OffPA2_Taylor_add}).

As in the case of LOLA-OffPA2, we can use an automatic differentiation engine to directly compute the gradients. Algorithm \ref{alg:LA-OffPA2} in Appendix illustrates the LA-OffPA2 optimization framework for the case of $n$ agents. At each state $s\sim\rho^b(s)$ the agent $i \in \mathcal{N}$ first anticipates the changes in actions of all agents $j \in \{\mathcal{N}-\{i\}\}$ using Eq. (\ref{eq:LOLA-OffPA2_add}). Then, agent $i$ updates its parameters $\theta_i$ by the following gradient adjustment:
\begin{equation}
\begin{split}
    \nabla_{\theta_i}J_i^{\text{LA-OffPA2}} = \mathbb{E}_{\rho^\beta(s)} \nabla_{\theta_i}\mu_{\theta_i}(s)\nabla_{a_i}Q_i(s,a_1+\perp\Delta a_1,...,a_i,...,a_n+\perp\Delta a_n)|_{a_i=\mu_{\theta_i}(s)\;\forall i \in \mathcal{N}}.
\end{split}
\end{equation}

\subsubsection{HLA-OffPA2}
As previously mentioned, HLA is proposed to improve coordination in fully cooperative games with common interested agents. To develop HLA-OffPA2, we first define $\mathcal{M} \subseteq \mathcal{N}$, as a set of size $m=|\mathcal{M}|$ common-interested agents with a common reward function $\mathcal{R}=\mathcal{R}_i=\mathcal{R}_j \;\forall i,j \in \mathcal{M}$. Without the loss of generality, we consider $\mathcal{M} = \mathcal{N}$, i.e., team games with a common state-action value function $Q(s,a_1,...,a_m)$. 

\textbf{Hierarchy level assignment.} 
Similarly to HLA, each agent in HLA-OffPA2 is first assigned to one of $m$ levels, with level one as the lowest hierarchy level and level $m$ as the highest. Although the hierarchy level assignment can be random as proposed by \cite{Bighashdel2023}, we utilize the amount of influence that agents have on others, i.e., their shaping capacity, as the indicator for the agents' hierarchy levels (see Section \ref{apsec:Ablation study on hierarchy-level assignments} for experimental comparisons). We define the shaping capacity of the $i^\text{th}$ agent, $\mathcal{SC}_i$, as the sum of the action shaping values with respect to all other agents $j$: 
\begin{equation}
\label{eq:hrmaddpg1}
\begin{split}
    \mathcal{SC}_i=\sum_{j \in \{\mathcal{M}-\{i\}\}}\norm{(\nabla_{a_i} \Delta a_j)^\intercal \nabla_{a_j}Q(a_1,...,a_m)},
\end{split}
\end{equation}
where $\Delta a_j=\nabla_{a_j}Q(a_1,...,a_m)$. The agent with the highest shaping capacity is assigned to the highest hierarchy level, and so on. As HLA-OffPA2 benefits from centralized learning, the only constraint for HLA reasoning rules remains the centralized state-action value function. 

\begin{figure}
    \centering
    \subfloat{\includegraphics[width=1\textwidth]{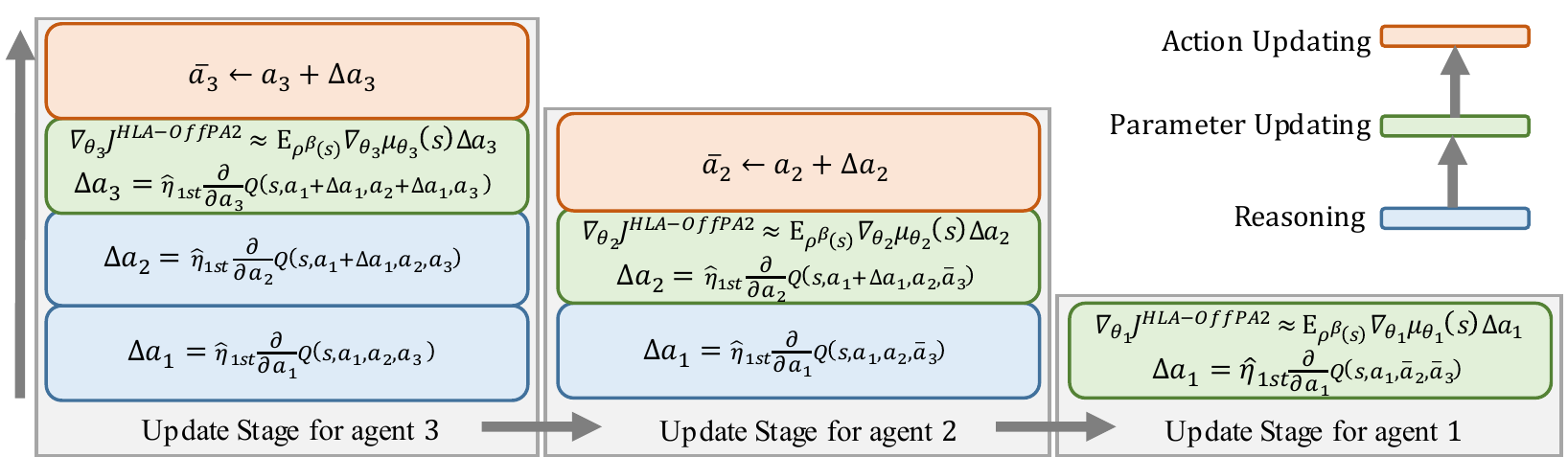}}%
    \captionof{figure}{\label{fig:hr} An example of the parameter update stages in HLA-OffPA2, for a game with three common-interested agents, where agent 1, agent 2, and agent 3 are assigned to hierarchy level 1, hierarchy level 2, and level 3, respectively.}
\end{figure}

\textbf{Update rules.} After the hierarchy level assignment, the agents update their policy parameters in $m$ update stages, i.e., one for each agent, and in a top-down fashion: the agent in the highest hierarchy level updates its policy parameters first. In each update stage, the corresponding agent 1) reasons about the actions of followers (if any) in a bottom-up fashion, i.e., it reasons about the agent in the lowest hierarchy level first, 2) updates its policy parameters, and 3) updates its action for the next update stage (if any). 

If we set $m=2$ and assume that agent Two is the leader (HLA-OffPA2-L) and agent One is the follower (HLA-OffPA2-F), the gradient adjustment for the leader is 
\begin{equation}
\label{eq:HLA-OffPA2-L_Taylor}
\begin{split}
    \nabla_{\theta_2}J^\text{HLA-OffPA2-L} &= \mathbb{E}_{\rho^\beta(s)} \nabla_{\theta_2}\mu_{\theta_2}(s)\nabla_{a_2}Q(s,a_1+\Delta a_1,a_2)|_{a_1=\mu_{\theta_1}(s),a_2=\mu_{\theta_2}(s)}\\
    & \approx \mathbb{E}_{\rho^\beta(s)} \nabla_{\theta_2}\mu_{\theta_2}(s)\left( \nabla_{a_2}Q + (\nabla_{a_1 a_2}Q)^\intercal \Delta a_1 + (\nabla_{a_2} \Delta a_1)^\intercal \nabla_{a_1}Q \right),
\end{split}
\end{equation}
where 
\begin{equation}
\label{eq:HLA-OffPA2-L_Taylor_add}
\begin{split}
    & \Delta a_1 = \hat{\eta}_{\text{1st}}\nabla_{a_1}Q(s,a_1,a_2)|_{a_1=\mu_{\theta_1}(s),a_2=\mu_{\theta_2}(s)}\\
    & Q = Q(s,a_1,a_2)|_{a_1=\mu_{\theta_1}(s),a_2=\mu_{\theta_2}(s)}.
\end{split}
\end{equation}
The shaping plan of the leader is to change its actions as
\begin{equation}
\begin{split}
    \bar{a}_2=a_2 + \hat{\eta}_{\text{1st}}\nabla_{a_2}Q(s,a_1+\Delta a_1,a_2)|_{a_1=\mu_{\theta_1}(s),a_2=\mu_{\theta_2}(s)},
\end{split}
\end{equation}
so that an optimal increase in the common state-action value is achieved after its new actions are taken into account by the follower. Therefore, the follower adjusts its parameters through 
\begin{equation}
\label{eq:HLA-OffPA2-F_Taylor}
\begin{split}
    \nabla_{\theta_1}J^\text{HLA-OffPA2-F} &= \mathbb{E}_{\rho^\beta(s)} \nabla_{\theta_1}\mu_{\theta_1}(s)\nabla_{a_1}Q(s,a_1,\bar{a}_2)|_{a_1=\mu_{\theta_1}(s)}\\
    & \approx \mathbb{E}_{\rho^\beta(s)} \nabla_{\theta_1}\mu_{\theta_1}(s)\left( \nabla_{a_1}Q + (\nabla_{a_1 a_2}Q)^\intercal \hat{\eta}_{\text{1st}}\nabla_{a_2}Q(s,a_1+\Delta a_1,a_2)\right).
\end{split}
\end{equation}

As in the case of LOLA-OffPA2 and LA-OffPA2, we can use an automatic differentiation engine to directly compute the gradients. To clarify the update rules in HLA-OffPA2 for $m>2$, we demonstrate an example of update stages for three common-interested agents in Figure \ref{fig:hr}, where agent One, agent Two, and agent Three are assigned to hierarchy level 1, hierarchy level 2, and hierarchy level 3, respectively. For the case of $m$ agents, see the HLA-OffPA2 optimization framework in Algorithm \ref{alg:hrlamaddpg} in Appendix.

\section{Experiments}
\label{sec:Experiments}
In this section, we conduct a set of experiments to accomplish two main goals: 1) to indicate the benefits of learning anticipation in a broader range of MARL problems, including non-differentiable games with large state spaces, and 2) to show the advantages of our proposed action anticipation approach with respect to policy parameter anticipation.

To accomplish the first goal, we compare our proposed methods with Multi-Agent Deep Deterministic Policy Gradient (MADDPG)~\citep{lowe2017multi}, configured with three state-of-the-art update rules: 1) standard update rule \citep{lowe2017multi}, referred to as MADDPG, 2) Centralized Policy Gradient (CPG) update rule \citep{Peng2021}, referred to as CPG-MADDPG, and 3) Probabilistic Recursive Reasoning (PR2) update rule \citep{wen2019probabilistic}, referred to as PR2-MADDPG. To achieve our second goal, we compare our proposed OffPA2-based methods with existing HOG methods that are capable of solving non-differentiable games (see Figure \ref{fig:taxonomy}). Specifically, we compare LOLA-OffPA2 and LA-OffPA2 with LOLA-DiCE and LA-DiCE, respectively. Prior work \citep{foerster2018dice} has shown that LOLA-DiCE significantly outperforms LOLA in IPD, and, consequently, we do not compare LOLA-OffPA2 with LOLA. As both LOLA-DiCE and LA-DiCE are based on policy parameter anticipation, with these experiments, we can highlight the benefits of our novel action anticipation approach. Similarly to the implementation of \cite{foerster2018dice}, agents in DiCE can access the policy parameters of other agents. Since the original HLA method can only be applied to differentiable games, we do not compare HLA-OffPA2 with HLA. 

Unless mentioned otherwise, we evaluate the performance of methods based on (normalized) Average Episode Reward (AER), with higher values indicating better performance. Furthermore, we assess the efficiency of HOG methods based on the Learning Anticipation Time Complexity (LATC), which for HOG method $\mathcal{H}$ is computed as:
\begin{equation}
\begin{split}
    \text{LATC}(\mathcal{H}) = \frac{\text{per iteration training time of }\mathcal{H}}{\text{per iteration training time of the na\"ive version of }\mathcal{H}} - 1 \geq 0
\end{split}
\end{equation}
where the na\"ive version of $\mathcal{H}$ does not perform learning anticipation. The lower values of LATC indicate better efficiency, and $\text{LATC}=0$ implies that learning anticipation adds zero time complexity to the algorithm. In the following sections, we separately evaluate our proposed methods and discuss the results (see Appendix \ref{apsec:Implementation details} for implementation details).

\subsection{Evaluation of LA-OffPA2}
\begin{figure}
  \begin{minipage}[c]{.5\textwidth}
    \vspace{0.1cm}
    \subfloat{\includegraphics[width=0.9\textwidth]{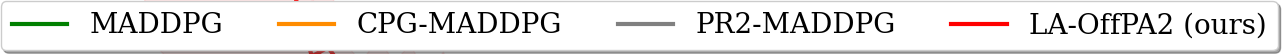}}\\[-2ex]
    \subfloat{\includegraphics[width=0.9\textwidth]{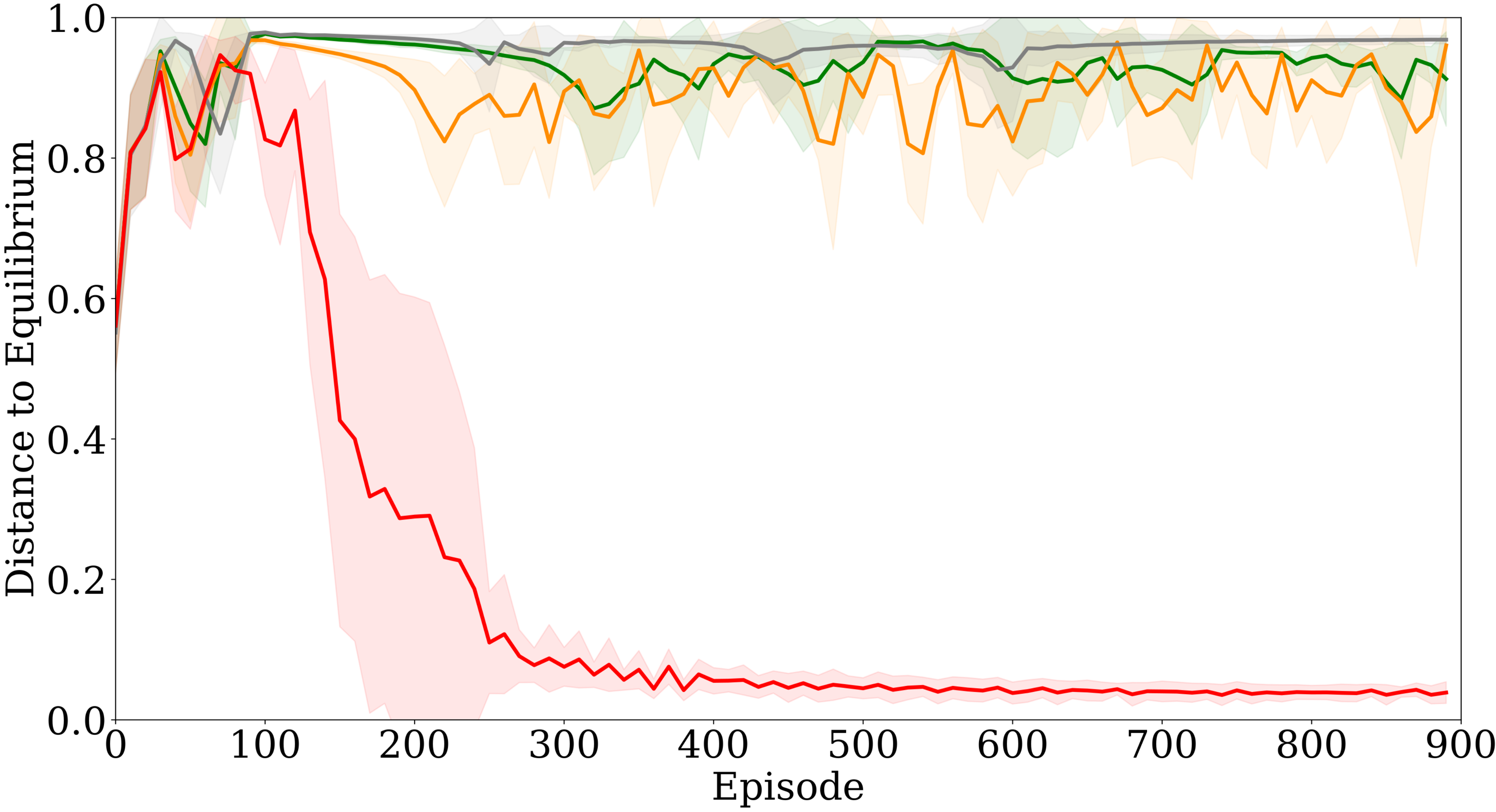}}
    \captionof{figure}{\label{fig:irg} Learning curves in iterated rotational game in terms of the distance to the equilibrium point ($\downarrow$).}
\end{minipage}
\hspace{0.2cm}
\begin{minipage}[c]{.45\linewidth}

  \vfill
  \vspace{1.2cm}
  \begin{adjustbox}{width=\linewidth}
    \begin{tabular}{lc }
    	Method   &
    	Distance to Equilibrium $\downarrow$ \\
    	\toprule
    	LA-DiCE & 0.09$\pm$0.07  \\
    	LA-OffPA2 (ours) & \textbf{0.03$\pm$0.02}  \\
        \bottomrule 
         & \\

        Method   &
    	Learning Anticipation Time Complexity $\downarrow$\\
    	\toprule
    	LA-DiCE   &  1.06  \\
    	LA-OffPA2 (ours) &  \textbf{0.13}  \\
    	\bottomrule
    \end{tabular}
    \end{adjustbox} 
    \vspace{1cm}
    \captionof{table}{\label{table:IRG} Comparison of LA principles in the frameworks of DiCE and our proposed OffPA2 in iterated rotational game.}
    
  \end{minipage}
\end{figure}

\begin{wrapfigure}{r}{0.45\textwidth}
\begin{minipage}[c]{1\linewidth}
  \vfill
  \begin{adjustbox}{width=1\linewidth}
    \begin{tabular}{c|c|c }
    	   & discrete action 1 & discrete action 2  \\
    	\toprule
    	discrete action 1 & $(0,3)$ & $(3,2)$  \\
    	\toprule
    	discrete action 2 & $(1,0)$ & $(2,1)$  \\
    \end{tabular}
    \end{adjustbox} 
    \captionof{table}{\label{table:IRG_rewards} Rewards in iterated rotational game.}  
\end{minipage}
\end{wrapfigure}
We evaluate the methods on the non-differentiable version of the rotational game proposed by \cite{zhang2010multi}, and we refer to it as the Iterated Rotational Game (IRG). IRG is a one-state, two-agent, one-action (continuous) matrix game with the rewards depicted in Table \ref{table:IRG_rewards} (for two discrete actions). However, the agents do not have access to the reward table and can only receive a reward for their joint actions. Each agent $i \in \{1,2\}$ must choose a 1-D continuous action ($0\leq a_i\leq 1$ representing the probability of taking two discrete actions. The game has a unique equilibrium point at $a_1=a_2=0.5$, which is also the fixed point of the game. The rotational game was originally proposed to demonstrate the circular behavior that can emerge if the agents follow the na\"ive gradient updates. LA agents, on the other hand, can quickly converge to the equilibrium point by considering their opponent's parameter adjustment. We evaluate the performances of methods based on the Distance to Equilibrium (DtE), which is the Euclidean distance between current actions and the equilibrium point.

Figure~\ref{fig:irg} demonstrates the learning curves for LA-OffPA2 and other, state-of-the-art MADDPG-based algorithms. From this figure, we find that LA-OffPA2 is the only method that can converge to equilibrium actions. These results highlight the importance of learning anticipation in IRG. To further show the effectiveness of LA-OffPA2, we compare our LA-OffPA2 method with LA-DiCE \citep{foerster2018dice} and report the results in Table \ref{table:IRG}. Looking at DtE results in Table \ref{table:IRG}, it is apparent that both methods can solve the games, with LA-OffPA2 achieving slightly better results. However, if we compare the methods regarding learning anticipation time complexity (LATC), it is clear that LA-OffPA2 is significantly more efficient than LA-DiCE, which indicates the benefits of our OffPA2 framework with respect to DiCE.

\subsection{Evaluation of LOLA-OffPA2}
\begin{figure}
  \begin{minipage}[c]{.5\textwidth}
    \vspace{0.1cm}
    \subfloat{\includegraphics[width=0.9\textwidth]{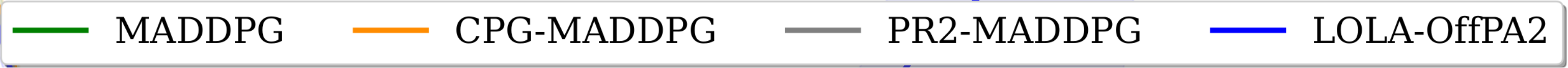}}\\[-2ex]
    \subfloat{\includegraphics[width=0.9\textwidth]{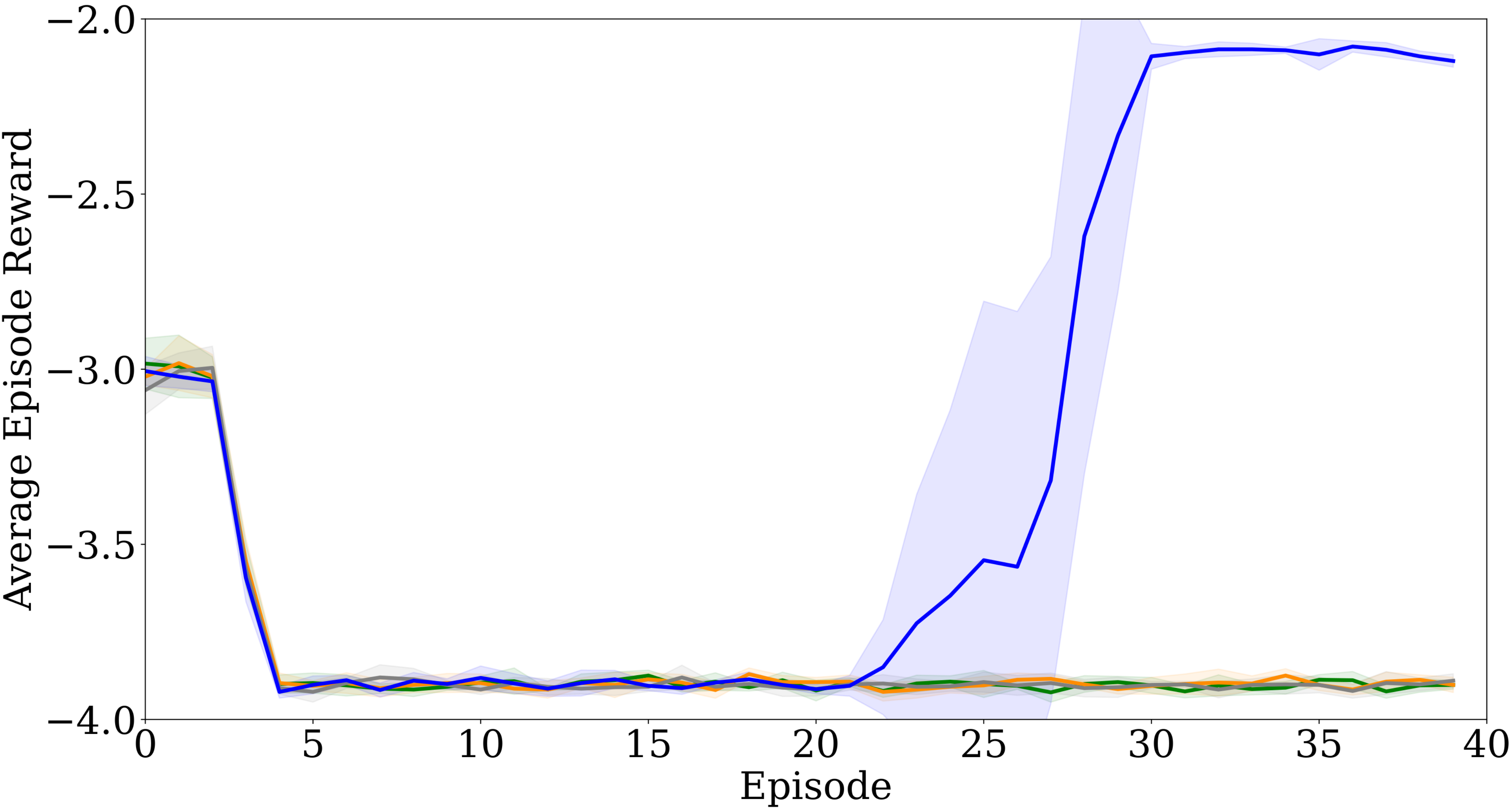}}
    \captionof{figure}{\label{fig:ipd} Learning curves in iterated prisoner's dilemma in terms of the average episode reward ($\uparrow$). }
\end{minipage}
\hspace{0.2cm}
\begin{minipage}[c]{.45\linewidth}

  \vfill
  \vspace{1.2cm}
  \begin{adjustbox}{width=\linewidth}
    \begin{tabular}{lc }
    	Method   &
    	Average Episode Reward $\uparrow$ \\
    	\toprule
    	LOLA-DiCE   &  -2.16$\pm$0.12  \\
    	LOLA-OffPA2 (ours) &  \textbf{-2.08$\pm$0.02}  \\
        \bottomrule 
         & \\

        Method   &
    	Learning Anticipation Time Complexity $\downarrow$\\
    	\toprule
    	LOLA-DiCE   &  1.18  \\
    	LOLA-OffPA2 (ours) &  \textbf{0.16}  \\
    	\bottomrule
    \end{tabular}
    \end{adjustbox} 
    \vspace{1cm}
    \captionof{table}{\label{table:IPD} Comparison of LOLA principles in the frameworks of DiCE and our proposed OffPA2 in iterated prisoner's dilemma.}
    
  \end{minipage}
\end{figure}

\subsubsection{Iterated prisoner's dilemma}
\label{sec:Matrix games}
\begin{wrapfigure}{r}{0.32\textwidth}
\begin{minipage}[c]{1\linewidth}
  \vfill
  \begin{adjustbox}{width=1\linewidth}
    \begin{tabular}{c|c|c }
    	   & Cooperate & Defect \\
    	\toprule
    	Cooperate & $(-1,-1)$ & $(-3,0)$  \\
    	\toprule
    	Defect & $(0,-3)$ & $(-3,-3)$  \\
    \end{tabular}
    \end{adjustbox} 
    \captionof{table}{\label{table:IPD_rewards} Rewards in iterated prisoner's dilemma.}  
\end{minipage}
\end{wrapfigure}
Iterated Prisoner's Dilemma (IPD) \citep{foerster2018learning} is a five-state, two-agent, two-action game with the reward matrices depicted in Table \ref{table:IPD_rewards}. Each agent must choose between two discrete actions (cooperate or defect). The game is played for 150 time steps ($T=150$). In the one-shot version of the game, there is only one Nash equilibrium for the agents (Defect, Defect). In the iterated games, (Defect, Defect) is also a Nash equilibrium. However, a better equilibrium is Tit-For-Tat (TFT), where the players start by cooperating and then repeat the previous action of the opponents. The LOLA agents can shape the opponent's learning to encourage cooperation and, therefore, converge to TFT \citep{letcher2018stable}. We evaluate the methods' performances based on the Averaged Episode Reward (AER).

In Figure~\ref{fig:ipd}, we depict the learning curves for LOLA-OffPA2 and the MADDPG-based methods. From this figure, we find that only LOLA-OffPA2 can solve the game, which once again highlights the importance of learning anticipation. Additionally, we compared the performance of LOLA-OffPA2 with LOLA-DiCE \citep{foerster2018dice}, which is designed specifically for this game, and we reported the results in Table~\ref{table:IPD}. Although both methods demonstrate high values of AER, our LOLA-OffPA2 is significantly more efficient as its LATC value is much lower than that of LOLA-DiCE.

\subsubsection{Multi-level Exit-Room game}
\label{sec:exit-room}
\begin{figure}[t]
  \centering
  \begin{minipage}[c]{0.85\textwidth}
    \centering
    \subfloat{\label{fig:exitroom_l1}\includegraphics[width=0.23\textwidth]{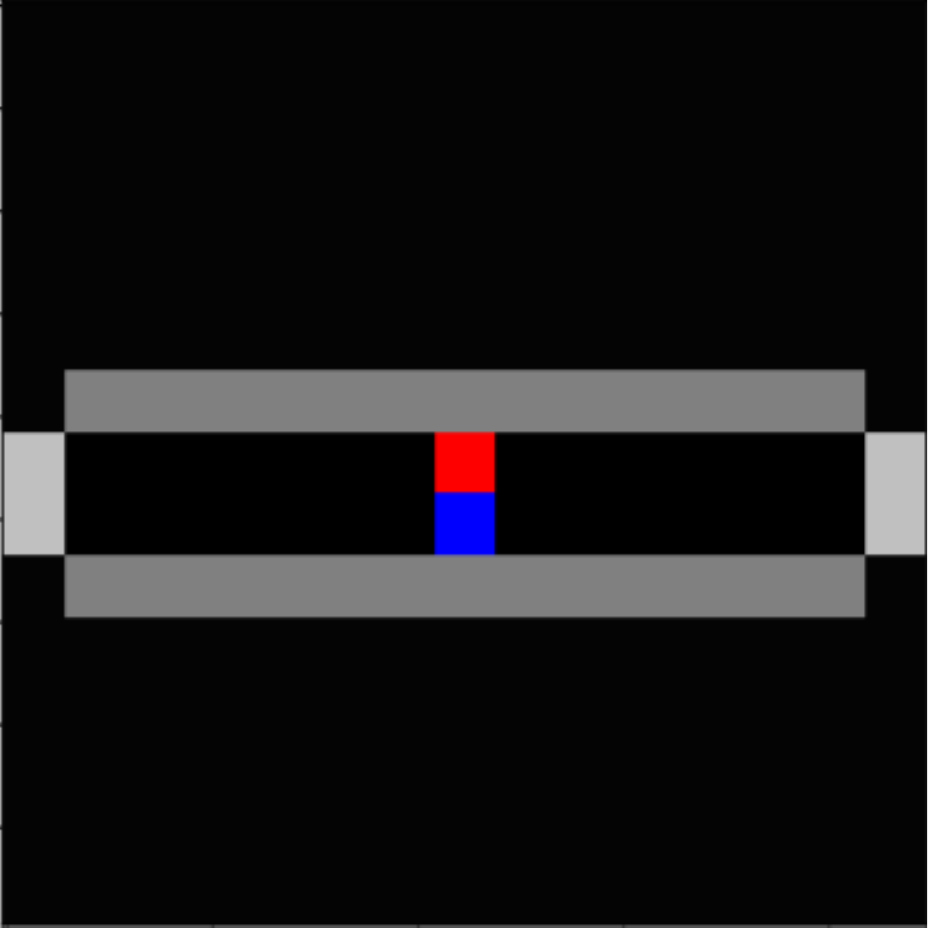}}%
    \hspace{1cm}\subfloat{\label{fig:exitroom_l2}\includegraphics[width=0.23\textwidth]{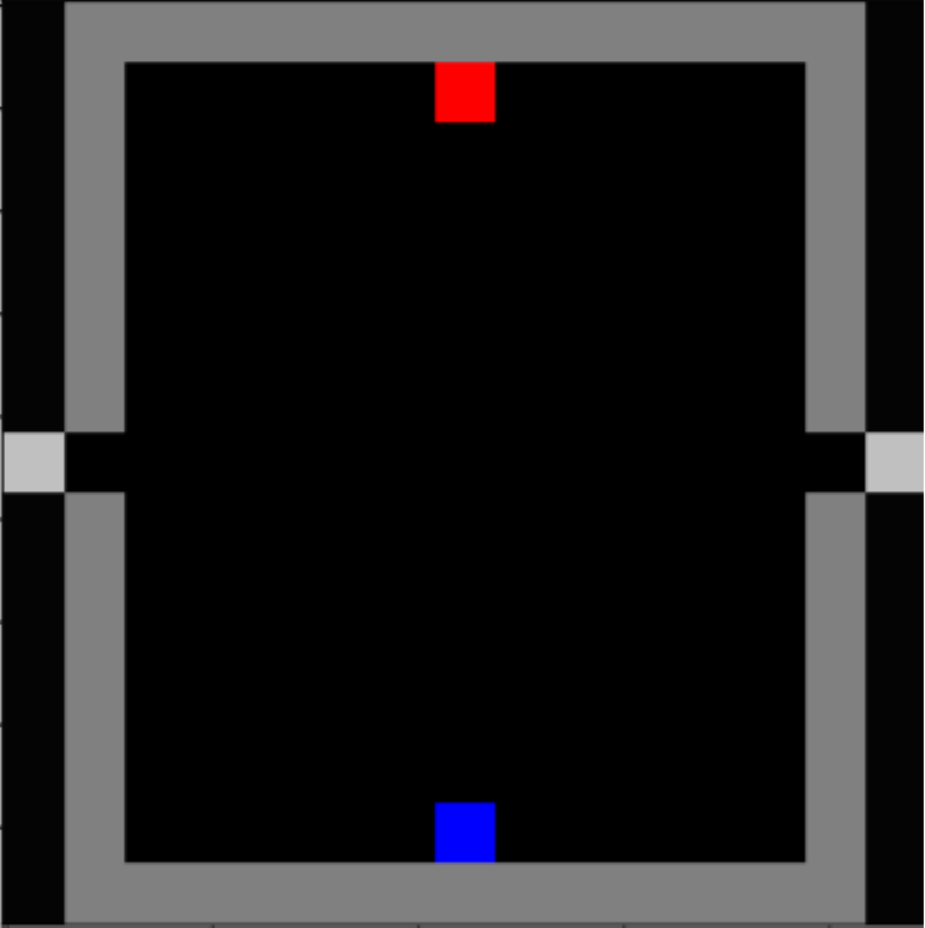}}%
    \hspace{1cm}\subfloat{\label{fig:exitroom_l3}\includegraphics[width=0.23\textwidth]{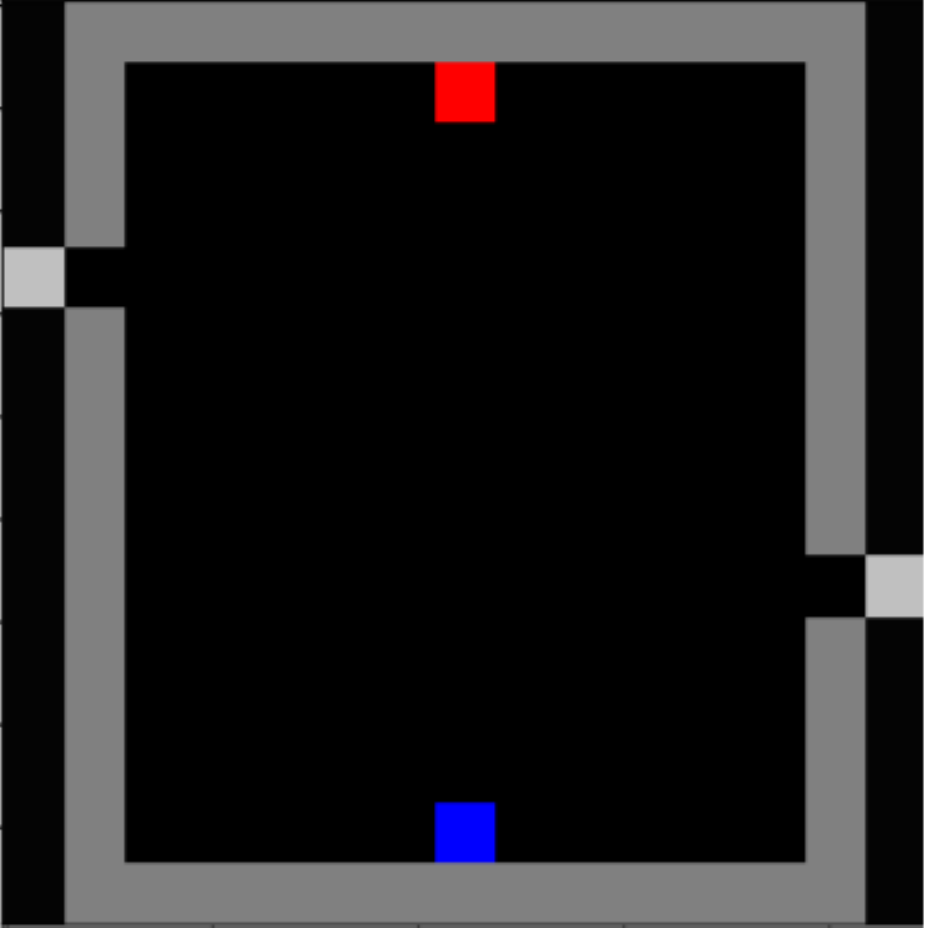}}
    \captionof{figure}{\label{fig:exitroom} State observation in the Exit-Room game, level one (left), level two (middle), and level three (right).}
\end{minipage}
\end{figure}

In the second experiment, we evaluate the capability of LOLA-OffPA2 in games with large state spaces, which is the envisioned use case for our framework. Inspired by \cite{SSDOpenSource}, we propose an Exit-Room game with three levels of complexity (see Figure \ref{fig:exitroom}). The Exit-Room game is a grid-world variant of the IPD, with two agents (blue and red) and $15^{2l}$ states where $l \in \{1,2,3\}$ is the complexity level of the game. The agents should cooperate and move toward the exit doors on the right. However, they are tempted to exit through the left doors and, in some cases, not exiting at all. In level 1, the agents have three possible actions (\textit{move-left}, \textit{move-right}, or \textit{do nothing}), and the reward is computed as \cite{SSDOpenSource}:
\begin{equation}
\label{eq:exit_room_reward}
\begin{split}
    &\text{reward}_{C} = \lambda_C(\text{cooperation}_{self} + \text{cooperation}_{opponent})\\
    &\text{reward}_{D} = \lambda_D(1-\text{cooperation}_{self})\\
    & \text{reward} = \text{reward}_{C} + \text{reward}_{D},
\end{split}
\end{equation}
where $\lambda_C$ and $\lambda_D$ are some constants, and $\text{cooperation}_{self}$ and $\text{cooperation}_{opponent}$ are the normalized distances of the agent and its opponent to the right door, respectively. In levels 2 and 3, the agents have additional \textit{move-up} and \textit{move-down} actions. In level 3, the door positions are randomly located, resulting in more complex interactions among the agents. In addition to the reward in Eq. (\ref{eq:exit_room_reward}), the agents receive an additional reward for approaching the doors in levels 2 and 3. Each agent receives four  $90\times90$ RGB images representing the state observations of the last four time steps.

\begin{figure}[t]
  \centering
  \begin{minipage}[c]{1\textwidth}
    \centering
    \subfloat{\includegraphics[width=0.5\textwidth]{Figures/IPDlegend.pdf}}\\[-2ex]
    \setcounter{subfigure}{0}
    \subfloat[Exit-Room level one.]{\label{fig:erl1}\includegraphics[width=0.33\textwidth]{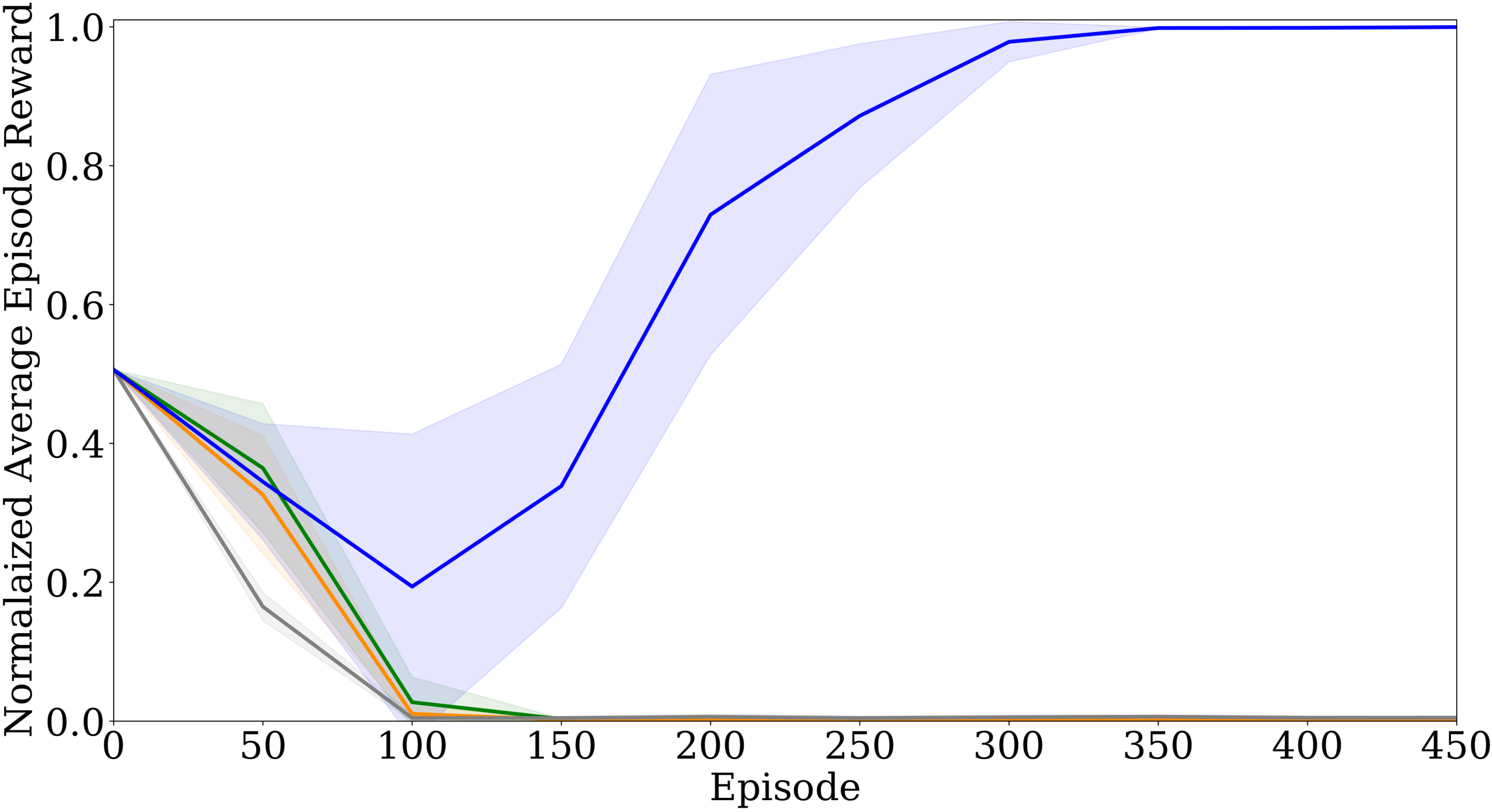}}%
    \subfloat[Exit-Room level two.]{\label{fig:erl2}\includegraphics[width=0.33\textwidth]{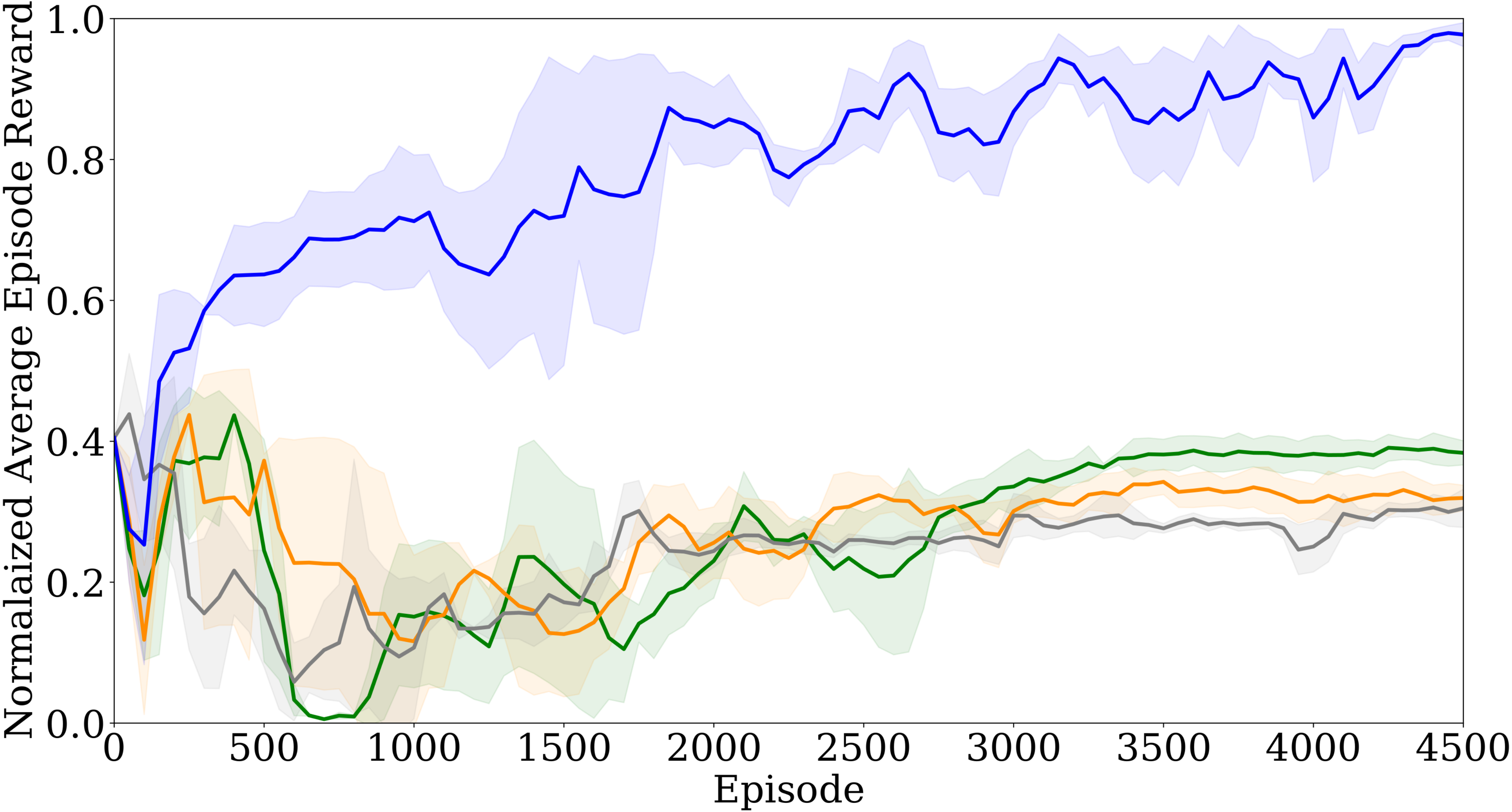}}%
    \subfloat[Exit-Room level three.]{\label{fig:erl3}\includegraphics[width=0.33\textwidth]{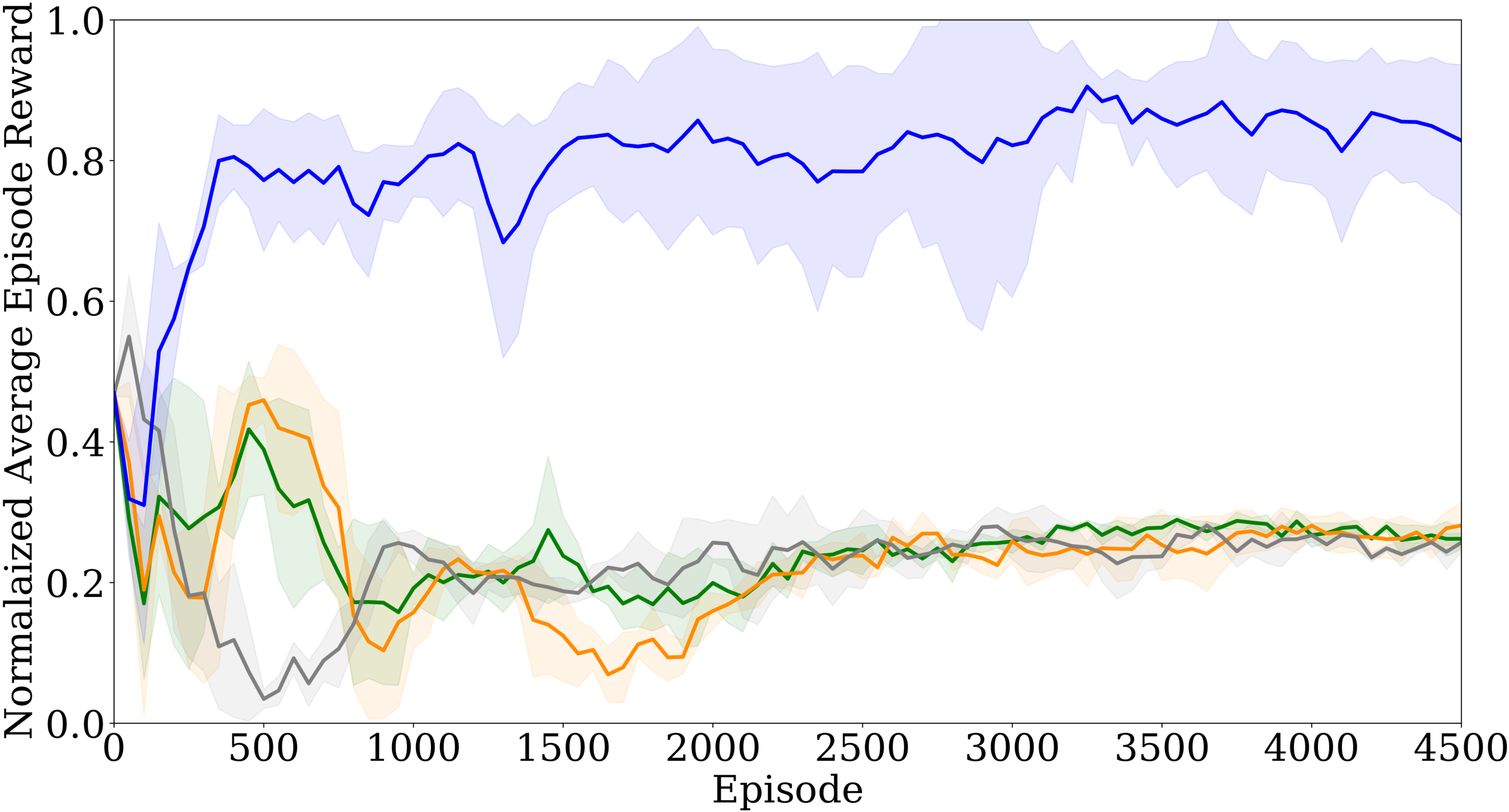}}
    \captionof{figure}{\label{fig:er} Learning curves in different complexity levels of the exit-room game in terms of the normalized average episode reward ($\uparrow$).}
\end{minipage}
\vspace{-\intextsep}
\end{figure}

\begin{table}[t]
\begin{minipage}{\linewidth}
  \begin{adjustbox}{width=\linewidth}
    \begin{tabular}{l|  ccc|ccccc }
    	\toprule
    	\multicolumn{1}{c}{}   &
    	\multicolumn{3}{c}{Normalized Average Episode Reward $\uparrow$}  &
    	\multicolumn{5}{c}{Learning Anticipation Time Complexity $\downarrow$}  \\
    	\cmidrule(lr){2-4}
    	\cmidrule(lr){5-9}
    	Methods     & $l=1$ & $l=2$ & $l=3$ & Na\"ive &1st-order & 2nd-order & 3rd-order & 4th-order\\
    	\toprule
    	LOLA-DiCE & 0.91$\pm$0.04 & 0.68$\pm$0.06 & 0.56$\pm$0.12 & 0.00 & 1.39 & 2.74 & 4.12 & 5.41 \\
    	LOLA-OffPA2 (ours) & \textbf{1.00$\pm$0.00} & \textbf{0.99$\pm$0.01} & \textbf{0.93$\pm$0.03} & 0.00 & \textbf{0.24} & \textbf{0.47} & \textbf{0.69} & \textbf{0.94} \\
    	\bottomrule
    \end{tabular}
    \end{adjustbox} 
    \captionof{table}{\label{table:er_dice} Comparisons of LOLA-DiCE with our proposed LOLA-OffPA2 in the Exit-Room game, in terms of performance (normalized average return in different game levels) and efficiency (learning anticipation time complexity in different reasoning levels).}
\end{minipage}
\end{table}

Figure \ref{fig:er} compares the learning curves of LOLA-OffPA2 and MADDPG-based methods in terms of Normalized Average Episode Reward (NAER), which is the AER value normalized between the highest and lowest episode rewards in each game level. In Figure \ref{fig:er}, we can clearly see that our LOLA-OffPA2 significantly outperforms the other methods, similarly in the IPD matrix game. 

To highlight the benefits of our proposed method with respect to existing HOG methods, we compare our LOLA-OffPA2 with LOLA-DiCE in terms of performance (by comparing NAER) and training efficiency (by comparing LATC) in Table~\ref{table:er_dice}. Observing Table \ref{table:er_dice}, it is apparent that LOLA-DiCE fails to acquire the highest rewards, particularly for the second and third levels of the game, where the state-space size is increased. The reason can be attributed to the fact that when the size of the state space increases, the LOLA-DiCE agents fail to properly approximate the higher-order gradients via sampling, and, consequently, cannot perform learning anticipation to achieve a higher reward. This is while our proposed LOLA-OffPA2 performs better in all levels of the game in terms of NAER, and scaling from na\"ive to higher-order reasoning is significantly more efficient for LOLA-OffPA2 than LOLA-DiCE. This emphasizes that we have overcome the limitations of HOG methods described in Section \ref{sec:Introduction}.

\subsubsection{Influence of the projected prediction length}
\label{sec:Influence of the projection estimation}
\begin{figure}[t]
	\centering
	\includegraphics[width=0.5\textwidth]{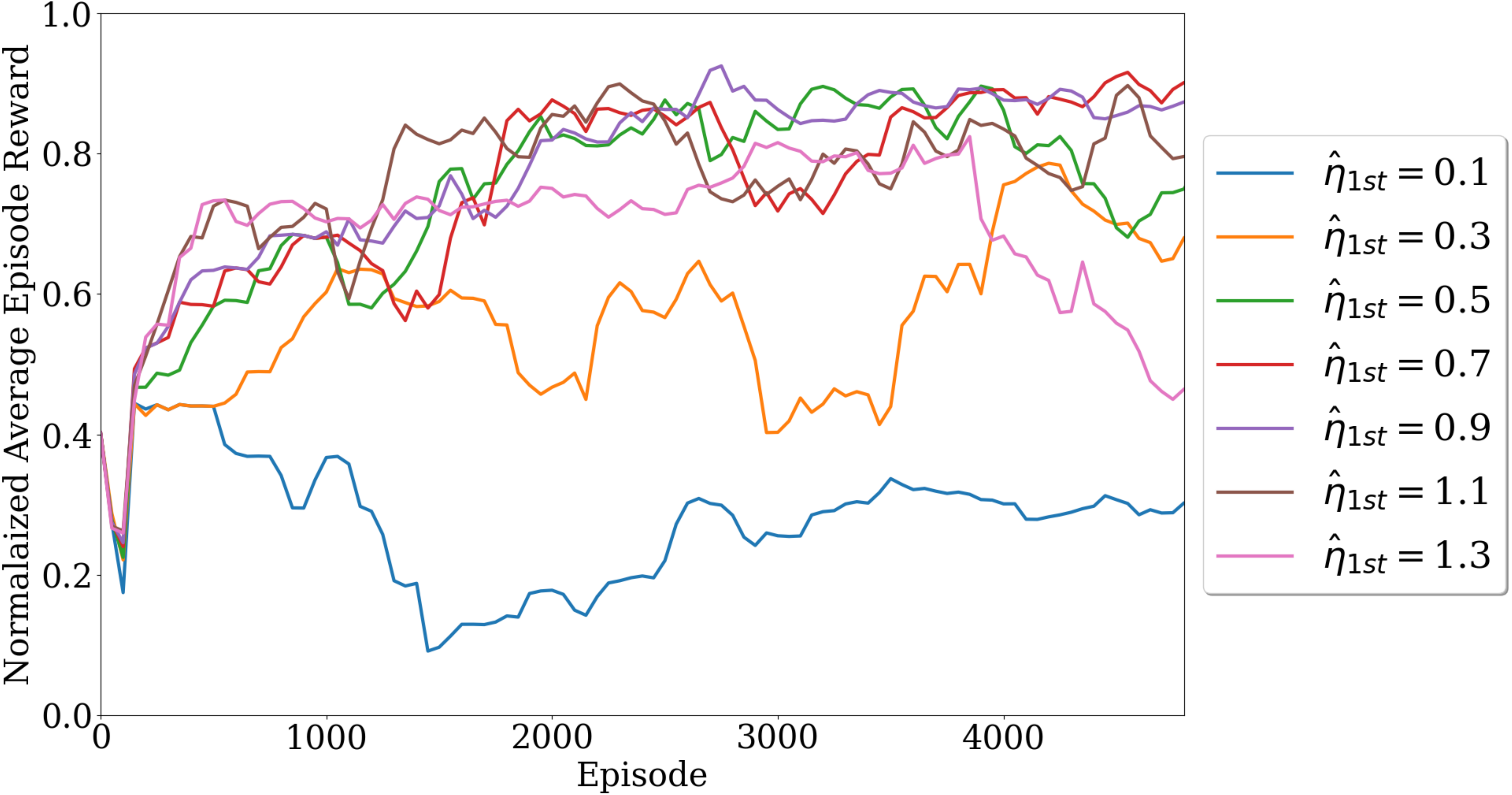}
	\caption{The influence of the projection prediction length ($\hat{\eta}_{\text{1st}}$) on the convergence behavior of LOLA-OffPA2 in the level-three Exit-room game.}
	\label{fig:projection}
\end{figure}
Based on our findings in Theorem \ref{th:projection_performance}, action anticipation via first-order Taylor expansion scales the prediction length and influences the convergence behavior of the HOG methods. Here, we empirically show that by directly changing the projected prediction length, i.e., $\hat{\eta}_{\text{1st}}$, we can tune the resulting prediction length in the state space, i.e., ${\eta}'$, and consequently improve the convergence behavior.
We conduct an experimental study to analyze the influence of $\hat{\eta}_{\text{1st}}$ on the convergence behavior of LOLA-OffPA2 in the level-three Exit-room game (See Figure \ref{fig:projection}). The experiments are repeated four times, and the mean results are reported in terms of NEAR in Figure \ref{fig:projection}. It is clear from Figure \ref{fig:projection} that by tuning $\hat{\eta}_{\text{1st}}$, we can alter the convergence behavior of LOLA-OffPA2. Furthermore, Figure \ref{fig:projection} demonstrates that low values of the projected prediction length ($\hat{\eta}_{\text{1st}}=0.1$) cancel the effect of learning anticipation in OffPA2, and high values of the projected prediction length ($\hat{\eta}_{\text{1st}}=1.3$) lead to instability of the LOLA-OffPA2 which can be attributed to our findings in Theorem \ref{th:projection_performance}.

\begin{figure}
  \begin{minipage}[c]{1\textwidth}
    \centering
    \subfloat[Game schematics]{\label{fig:sp_mis_game}\includegraphics[width=0.3\textwidth]{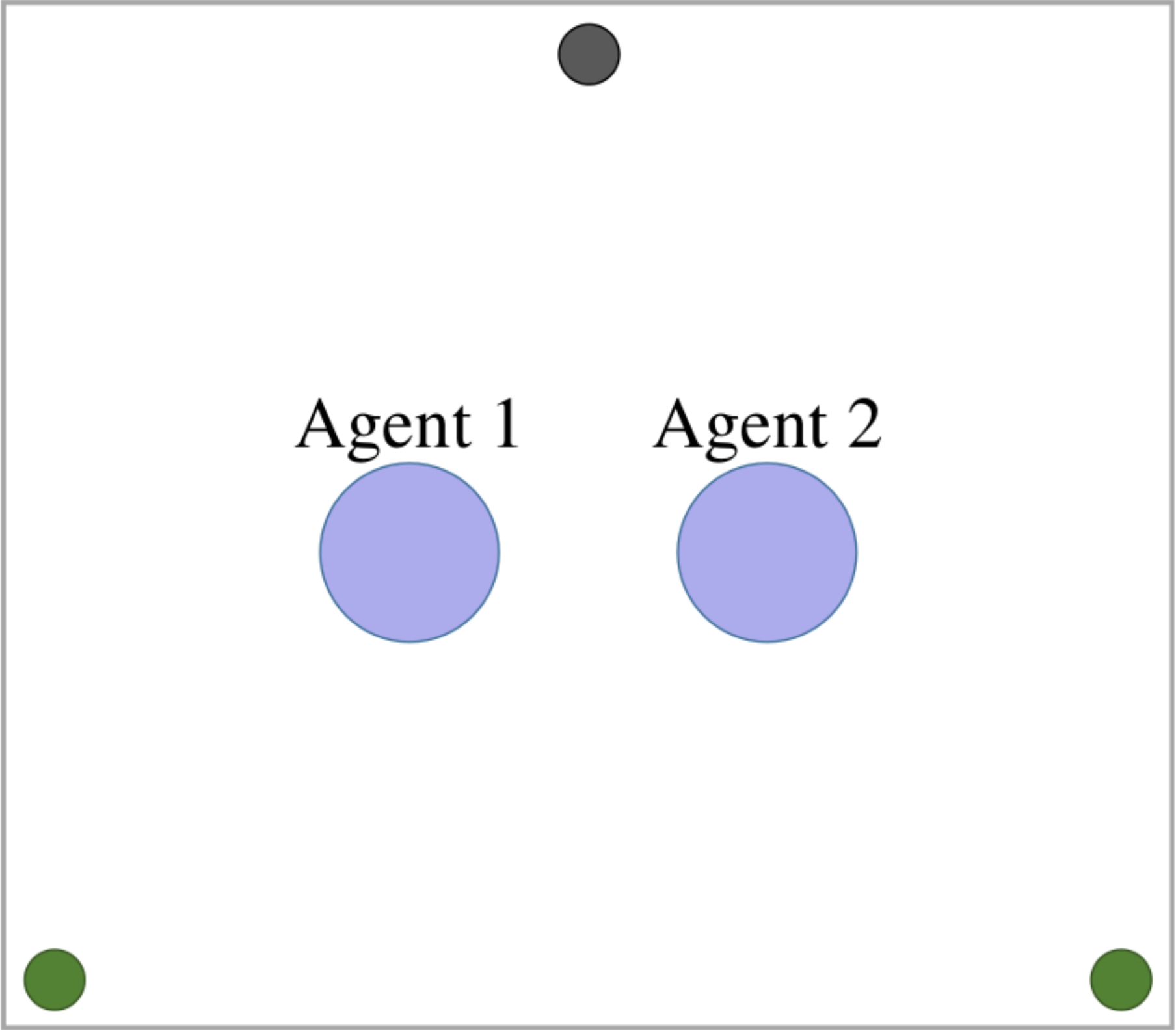}}
    \hspace{1cm}
    \subfloat[Learning curves]{\label{fig:sp_mis}\includegraphics[width=0.48\textwidth]{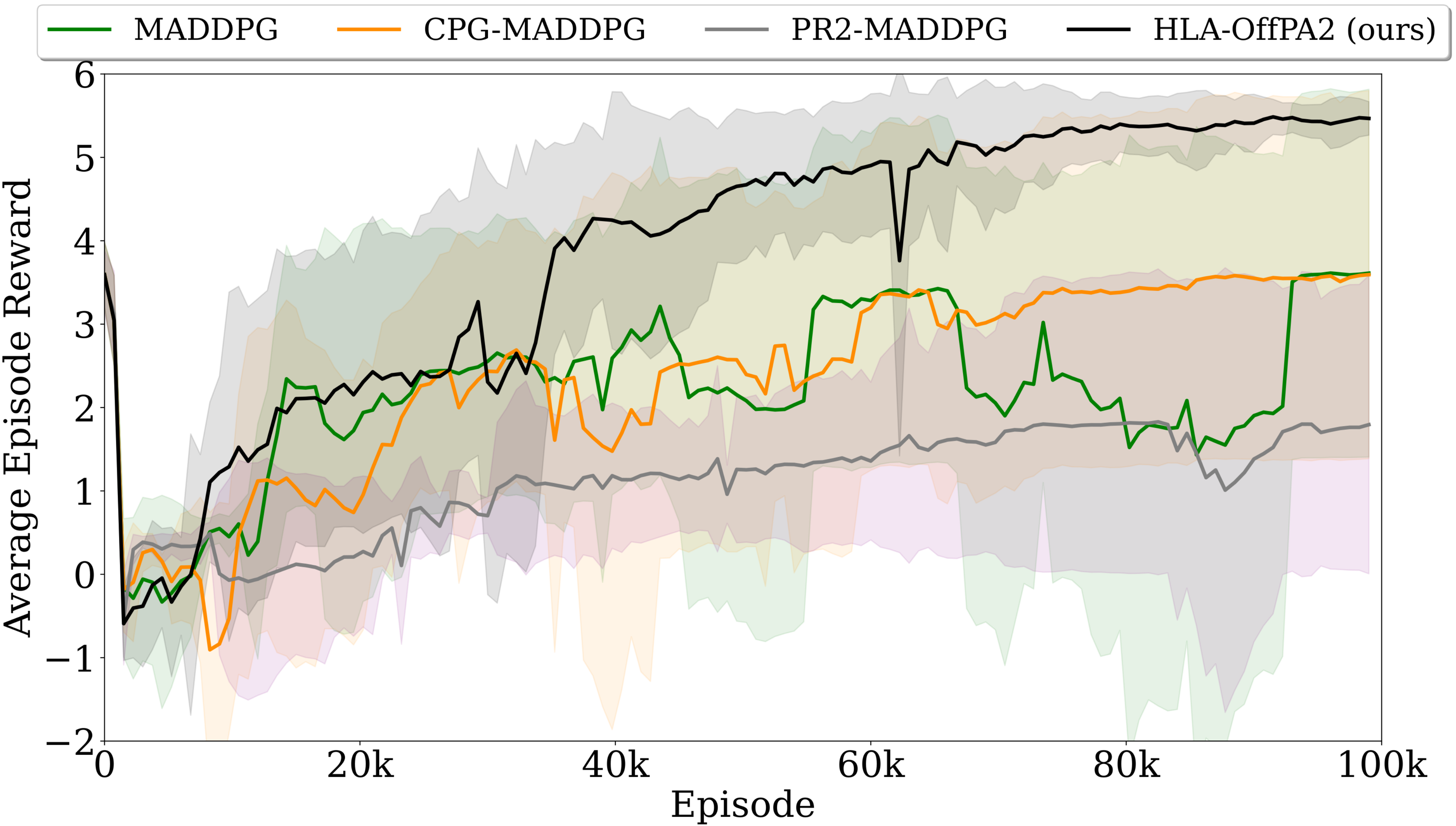}}
    \captionof{figure}{\label{fig:particle-coordination} Particle-coordination game. (a) Schematics of the game with two agents, i.e., purple circles, and three landmarks, i.e., gray and green circles. (b) Learning curves in terms of average episode reward ($\uparrow$). Best viewed in color.}
\end{minipage}
\end{figure}

\subsection{Evaluation of HLA-OffPA2}
\subsubsection{Particle-coordination game}
\begin{wrapfigure}{r}{0.35\textwidth}
\begin{minipage}[c]{1\linewidth}
  \vfill
  \begin{adjustbox}{width=1\linewidth}
    \begin{tabular}{c|c|c|c }
    	   & Green (L) & Gray & Green (R) \\
    	\toprule
    	Green (L) & 2 & 0 & -20  \\
    	\toprule
    	Gray & 0 & 0.4 & 0  \\
        \toprule
    	Green (R) & -20 & 0 & 2  \\
    \end{tabular}
    \end{adjustbox} 
    \captionof{table}{\label{table:sp_mis_rewards} Rewards in particle-coordination game.}  
\end{minipage}
\end{wrapfigure}
To demonstrate the coordination capability of HLA-OffPA2, we propose the Particle-Coordination Game (PCG) in the Particle environment \citep{lowe2017multi}. As shown in Figure \ref{fig:sp_mis_game}, each one of the two agents (purple circles) should select and approach one of the three landmarks (one gray and two green circles). Landmarks are selected based on the closest distance between the agent and the landmarks. Suppose the agents select and approach the same landmark. In that case, they receive global (by selecting the green landmarks) or local (by selecting the gray landmark) optimal rewards. They will receive an assigned miscoordination penalty if they select and approach different landmarks (see Table \ref{table:sp_mis_rewards}). Each agent receives a 10-D state observation vector (velocity and position information of the agent, i.e., 4-D, and location information of the landmarks, i.e., 6-D) and selects a 5-D, one-hot vector representing one of the five discrete actions: \textit{move-right}, \textit{move-left}, \textit{move-up}, \textit{move-down}, and \textit{stay}. The horizon is set to 25. The game is quite challenging as the agents cannot see the locations of each other, and they can be subject to miscoordination. 

In Figure~\ref{fig:sp_mis}, we depict the learning curves for our HLA-OffPA2 and other MADDPG-based methods in terms of the average episode reward (AER). As demonstrated, MADDPG-based methods have relatively high variance in the convergence points. For instance, the MADDPG agents, which heavily benefit from exploration and randomness during policy parameter updates, can occasionally converge to the global optimal point with the highest AER. However, the high miscoordination penalty forces the agents to choose the safest option (gray landmark), which leads to a zero reward in the worst-case scenario. From this figure, it is clear that our HLA-OffPA2 is the only method that consistently converges to the global optimum of the game, which is consistent with the reported results for HLA in fully-cooperative differentiable games \citep{Bighashdel2023}. 

\begin{table}
\begin{minipage}{\linewidth}
  \begin{adjustbox}{width=\linewidth}
    \begin{tabular}{l|  ccc|ccc }
    	\toprule
    	\multicolumn{1}{c}{}   &
    	\multicolumn{3}{c}{Particle Environment}  &
    	\multicolumn{3}{c}{Mujoco Environment}  \\
    	\cmidrule(lr){2-4}
    	\cmidrule(lr){5-7}
    	     & Cooperative Navigation & Physical Deception & Predator-Prey & Half-Cheetah & Walker & Reacher\\
    	\toprule
        Observation & 18-D    & 10-D (8-D) & 14-D (12-D) & 11-D & 11-D & 8-D \\
        Action      & 5-D     & 5-D (5-D) & 5-D (5-D) & 3-D & 3-D & 1-D \\
        Action type& discrete & discrete & discrete & continuous & continuous& continuous \\
        Horizon (step) & 25 & 25 & 25 & 100 & 300 & 50 \\       
    	\bottomrule
    \end{tabular}
    \end{adjustbox} 
    \captionof{table}{\label{table:dimensions} Specifications in the standard multi-agent games. In the mixed environments, the dimensions are reported as "$d_1$ $(d_2)$" where $d_1$ is the dimension for common-interested agents and $d_2$ is the dimension for self-interested ones.}
\end{minipage}
\end{table}

\begin{table}
\begin{minipage}{\linewidth}
  \begin{adjustbox}{width=\linewidth}
    \begin{tabular}{l|  ccc|ccc }
    	\toprule
    	\multicolumn{1}{c}{}   &
    	\multicolumn{3}{c}{$\uparrow$NAER in Particle Environment}  &
    	\multicolumn{3}{c}{$\uparrow$NAER in Mujoco Environment}  \\
    	\cmidrule(lr){2-4}
    	\cmidrule(lr){5-7}
    	Methods     & Cooperative Navigation & Physical Deception & Predator-Prey & Half-Cheetah & Walker & Reacher\\
    	\toprule
        DDPG (LB)       & 0.00 & 0.00 & 0.00 & 0.00 & 0.00 & 0.00 \\
        C-MADDPG (UB)   & 1.00 & 1.00 & 1.00 & 1.00 & 1.00 & 1.00 \\
    	\toprule
        MADDPG          & 0.77 & 0.61 & 0.21 & 0.86 & 0.45 & 0.02 \\
        CPG-MADDPG      & 0.78 & 0.67 & 0.18 & 0.88 & 0.46 & 0.05 \\
        PR2-MADDPG      & 0.78 & 0.54 & 0.08 & 0.85 & 0.45 & 0.01 \\
        HLA-OffPA2 (ours) & \textbf{0.88} & \textbf{0.83} & \textbf{0.44} & \textbf{0.94} & \textbf{0.67} & \textbf{0.42} \\
    	\bottomrule
    \end{tabular}
    \end{adjustbox} 
    \captionof{table}{\label{table:standard_games} Comparisons of methods in terms of the Normalized Average Episode Reward (NAER) for common-interested agents. LB: Lower Bound. UB: Upper Bound.}
\end{minipage}
\end{table}

\subsubsection{Standard multi-agent games}
For the final experiments, we evaluate our HLA-OffPA2 and MADDPG-based methods in three Particle environment games \citep{lowe2017multi}: 1) Cooperative Navigation with three common-interested agents, 2) Physical Deception with two common-interested agents and one self-interested agent, and 3) Predator-Prey with two common-interested (predator) agents and one self-interested (prey) agent. Furthermore, we compare the methods in three games within the multi-agent Mujoco environment \citep{Peng2021}: 1) two-agent Half-Cheetah, 2) two-agent Walker, and 3) two-agent Reacher. In the mixed environments (Physical Deception and Predator-Prey), we have employed the MADDPG method for the self-interested agents in all experiments. Games' specifications are reported in Table \ref{table:dimensions}. We created separate validation and test sets for each game that included 100 and 300 randomly generated scenarios, respectively. In each game, we save the models that have the best performance on the validation set and test them on the test set to report the results. All experiments are repeated five times, and the mean results are reported in Table \ref{table:standard_games} in terms of the Normalized AER (NAER). The normalization is done between the single-agent variant of MADDPG, i.e., DDPG \citep{lillicrap2016continuous}, and a fully centralized (in learning and execution) variant of MADDPG, referred to as C-MADDPG. As all of the games are non-differentiable, the original HLA method is no longer applicable.

In Table \ref{table:standard_games}, we observe that our proposed HLA-OffPA2 consistently and significantly outperforms all the state-of-the-art MADDPG-based methods. Again, these results confirm that learning anticipation, and in particular, our proposed HLA-OffPA2 improves coordination among common-interested agents, leading to better results. 

\subsubsection{Ablation study on hierarchy-level assignments}
\label{apsec:Ablation study on hierarchy-level assignments}
We have additionally conducted an ablation study on the hierarchy-level assignment in the HLA-OffPA2. Rather than iteratively sorting the agents based on their shaping capacities through Eq. (\ref{eq:hrmaddpg1}), we randomly assigned the agents to hierarchy levels in the beginning and fixed the hierarchy levels throughout the optimization. This variant of the HLA-OffPA2, i.e., referred to as HLA-OffPA2 (F), is evaluated and compared in Table \ref{table:standard_games_ablation}. As can be seen, using the proposed sorting strategy based on the shaping capacities of the agents, as done in our HLA-OffPA2, constantly improves performance. 

 \begin{table}[t]
\begin{minipage}{\linewidth}
  \begin{adjustbox}{width=\linewidth}
    \begin{tabular}{l|  ccc|ccc }
    	\toprule
    	\multicolumn{1}{c}{}   &
    	\multicolumn{3}{c}{$\uparrow$NAER in Particle Environment}  &
    	\multicolumn{3}{c}{$\uparrow$NAER in Mujoco Environment}  \\
    	\cmidrule(lr){2-4}
    	\cmidrule(lr){5-7}
    	Methods     & Cooperative Navigation & Physical Deception & Predator-Prey & Half-Cheetah & Walker & Reacher\\
    	\toprule
        HLA-OffPA2 (F)     & 0.85 & 0.80 & 0.38 & 0.92 & 0.63 & 0.38 \\
        HLA-OffPA2 & \textbf{0.88} & \textbf{0.83} & \textbf{0.44} & \textbf{0.94} & \textbf{0.67} & \textbf{0.42} \\
    	\bottomrule
    \end{tabular}
    \end{adjustbox} 
    \captionof{table}{\label{table:standard_games_ablation} Ablation study on the hierarchy level assignments in our HLA-OffPA2 method.}
\end{minipage}
\end{table}

 \section{Conclusion}
In this paper, we proposed the OffPA2 framework that enables the applicability of HOG methods to non-differentiable games with large state spaces. To indicate the advantages of our framework, we developed three novel HOG methods, LA-OffPA2, LOLA-OffPA2, and HLA-OffPA2. By conducting several experiments, we demonstrated that our proposed methods outperform the existing HOG methods in terms of performance and efficiency. Furthermore, we extensively compared our methods with various DPG-based methods, which do not use learning anticipation, and we signified that learning anticipation improves coordination among agents and leads to higher rewards. As a result of our framework,
the benefits of learning anticipation can now be used in many more MARL problems.





\newpage

\appendix
\section*{Appendix}
\section{Proofs}

\subsection{Proof of Theorem \ref{prop:action anticipation}}
\label{apsec:Proof of Proposition action}

At each state $s \sim \rho^b(s)$, agent One anticipates the changes in the policy parameters of agent Two, i.e., $\Delta \theta_2 (s)= \eta \nabla_{\theta_2} Q_2(s,\mu_{\theta_1}(s),\mu_{\theta_2}(s))$, and updates the policy parameters in the direction of: 
\begin{equation}
\label{eq:prop1}
\begin{split}
    \nabla_{\theta_1}J_1 = \mathbb{E}_{\rho^\beta(s)} \nabla_{\theta_1}Q_1(s,\mu_{\theta_1},\mu_{\theta_2+\Delta \theta_2(s)}(s)),
\end{split}
\end{equation}
In order to prove Proposition \ref{prop:action anticipation}, we need two assumptions: 1) neglecting the direct dependencies of the state-action value function on the policy parameters, and 2) first-order Taylor expansion. The first assumption is standard in off-policy reinforcement learning, including both deterministic and stochastic off-policy Actor-Critic algorithms \citep{silver2014deterministic, degris2012off}, and justification to support this assumption is provided in \cite{degris2012off}. As to the second assumption, please refer to Theorem \ref{th:projection_performance} to see how this assumption influences the performance.

Given the first assumption, we can rewrite Eq. (\ref{eq:prop1}) as 
\begin{equation}
\begin{split}
    \nabla_{\theta_1}J_1 = \mathbb{E}_{\rho^\beta(s)} \nabla_{\theta_1}\mu_{\theta_1}(s) Q_1(s,a_1,\Tilde{a}_2)|_{a_1=\mu_{\theta_1}(s),\Tilde{a}_2=\mu_{\theta_2+\Delta \theta_2(s)}(s)},
\end{split}
\end{equation}
Similarly, we can set $\Delta \theta_2 (s)= \eta \nabla_{\theta_2}\mu_{\theta_2}(s) \nabla_{a_2} Q_2(s,a_1,a_2)|_{a_2=\mu_{\theta_2}(s)}$. We now use the first-order Taylor expansion to map the anticipated gradient information to the action space:
\begin{equation}
\begin{split}
    \tilde{a}_2 &= \mu_{\theta_2+\Delta\theta_2(s)}(s)\\
    & \approx \mu_{\theta_2}(s) +\left(\Delta\theta_2(s)\right)^\intercal\nabla_{\theta_2}\mu_{\theta_2}(s) ,
\end{split}
\end{equation}
Given the definition of $\Delta \theta_2(s)$, we have:
\begin{equation}
\begin{split}
    \tilde{a}_2 & \approx \mu_{\theta_2}(s) +\left( 
    \eta \nabla_{\theta_2}\mu_{\theta_2}(s) \left(\nabla_{d_2} Q_2(s,a_1,a_2)\right)^\intercal\right)^\intercal\nabla_{\theta_2}\mu_{\theta_2}(s)\\
    & = \mu_{\theta_2}(s) +\nabla_{a_2}Q_2(s,a_1,a_2)  \left(\eta\nabla_{\theta_2}\mu_{\theta_2}(s)\right)^\intercal \nabla_{\theta_2}\mu_{\theta_2}(s)\\
    & = \mu_{\theta_2}(s) +\nabla_{a_2}Q_2(s,a_1,a_2) \eta \norm{\nabla_{\theta_2}\mu_{\theta_2(s)}}^2\\
    & = \mu_{\theta_2}(s) +\nabla_{a_2}Q_2(s,a_1,a_2)\hat{\eta}_{\text{1st}},
\end{split}
\end{equation}
where $\norm{.}$ is the $l^2$-norm and we have defined the projected prediction length $\hat{\eta}_{\text{1st}}=\eta \norm{\nabla_{\theta_2}\mu_{\theta_2(s)}}^2$, since $\norm{\nabla_{\theta_2}\mu_{\theta_2(s)}}^2$ is a positive number and independent of $\theta_1$. Therefore:
\begin{equation}
\label{eq:prop2}
\begin{split}
    \tilde{a}_2 &\approx a_2 +\Delta a_2.
\end{split}
\end{equation}
where we have defined $\Delta a_2=\hat{\eta}_{\text{1st}}\nabla_{a_2}Q_2(s,a_1,a_2)$
Replacing Eq. \ref{eq:prop2} in Eq. \ref{eq:prop1} yields:
\begin{equation}
\begin{split}
    \nabla_{\theta_1}J_1^{\text{LA}} \approx \mathbb{E}_{\rho^\beta(s)} \nabla_{\theta_1}\mu_{\theta_1}(s)\nabla_{a_1}Q_1(s,a_1,a_2+\Delta a_2)|_{a_1=\mu_{\theta_1}(s),a_2=\mu_{\theta_2}(s)},
\end{split}
\end{equation}
and consequently, Theorem \ref{prop:action anticipation} is proved.

\subsection{Proof of Theorem \ref{th:projection_performance}}
\label{apsec:Proof of Theorem projection}

In order to prove Theorem \ref{th:projection_performance}, we first need to show that:

\begin{lemma}
\label{lem:projection}
    If the anticipated changes are mapped from the policy parameter space to the action space using full-order Taylor expansion, there exists $\hat{\eta}_{\text{full}}\in \mathbb{R}$ such that 
\begin{equation}
\begin{split}
     \mu_{{\theta}_2+\Delta{\theta}_2}(s) = \mu_{{\theta}_2}(s) + \hat{\eta}_{\text{full}}\nabla_{a_2}Q_2(s,a_1,a_2),
\end{split}
\end{equation}
where
\begin{equation}
\begin{split}
     &\Delta\theta_2 =\eta \nabla_{\theta_2}\mu_{\theta_2}(s)\nabla_{a_2}Q_2(s,a_1,a_2)\\
     & \eta \in \mathbb{R}^+
\end{split}
\end{equation}
\end{lemma}
{\bf Proof}. The full-order Taylor expansion of the anticipated gradient yields:
\begin{equation}
\begin{split}
    \mu_{\theta_2+\Delta\theta_2}(s) & = \mu_{\theta_2}(s) +(\Delta\theta_2)^\intercal\nabla_{\theta_2}\mu_{\theta_2}(s) + \frac{1}{2}(\Delta\theta_2)^\intercal H_{\mu_{\theta_2}}(s)\Delta\theta_2+ O(\norm{\Delta\theta_2}^3),
\end{split}
\end{equation}
where $H_{\mu_{\theta_2}}(s)$ denotes the Hessian of $\mu_{\theta_2}$ at $s$. Given that:
\begin{equation}
\begin{split}
    \Delta\theta_2 =\left(\eta \nabla_{\theta_2}\mu_{\theta_2}(s)\nabla_{a_2}Q_2(s,a_1,a_2)\right)^\intercal,
\end{split}
\end{equation}
we have
\begin{equation}
\begin{split}
    \mu_{\theta_2+\Delta\theta_2}(s) = &  \mu_{\theta_2}(s) +\nabla_{a_2}Q_2(s,a_1,a_2) \eta\norm{\nabla_{\theta_2}\mu_{\theta_2(s)}}^2 \\
    &+ \frac{1}{2}\nabla_{a_2}Q_2(s,a_1,a_2) \eta^2\left(\nabla_{\theta_2}\mu_{\theta_2}(s) \right)^\intercal H_{\mu_{\theta_2}}(s) \nabla_{\theta_2}\mu_{\theta_2}(s)\left(\nabla_{a_2}Q_2(s,a_1,a_2)\right)^\intercal\\
    &+ O(\eta^3).
\end{split}
\end{equation}
By defining
\begin{equation}
\begin{split}
    &C1(s) = \norm{\nabla_{\theta_2}\mu_{\theta_2(s)}}^2 \\
    &C_2(s) = \frac{1}{2} \left(\nabla_{\theta_2}\mu_{\theta_2}(s) \right)^\intercal H_{\mu_{\theta_2}}(s)\nabla_{\theta_2}\mu_{\theta_2}(s)\left(\nabla_{a_2}Q_2(s,a_1,a_2)\right)^\intercal,
\end{split}
\end{equation}
we have:
\begin{equation}
\label{eq:V2}
\begin{split}
    \mu_{\theta_2+\Delta\theta_2}(s) = &  \mu_{\theta_2}(s) +\nabla_{a_2}Q_2(s,a_1,a_2) (\eta C_1(s) + \eta^2 C_2(s) + O(\eta^3)),
\end{split}
\end{equation}
Given the definition of $C_2(s)$ and the dimension constraint implied by $\nabla_{a_2}Q_2(s,a_1,a_2)$, it can be concluded that $C_1(s) \in \mathbb{R}^+$ and $C_2(s) \in \mathbb{R}$. Therefore:
\begin{equation}
\begin{split}
    \mu_{\theta_2+\Delta\theta_2}(s) = &  \mu_{\theta_2}(s) + \hat{\eta}_{\text{full}} \nabla_{a_2}Q_2(s,a_1,a_2) ,
\end{split}
\end{equation}
where $\hat{\eta}_{\text{full}}=\eta C_1(s) + \eta^2 C_2(s) + O(\eta^3)\in \mathbb{R}$. Consequently, we have proved Lemma \ref{lem:projection}

If we now map the anticipated changes of policy parameters, with a prediction length $\eta'\in \mathbb{R}^+$, to the action space using full-order Taylor expansion, we have: 
\begin{equation}
\begin{split}
     \mu_{{\theta}_2+\Delta{\theta'}_2}(s) = \mu_{{\theta}_2}(s) + \hat{\eta'}_{\text{full}}\nabla_{a_2}Q_2(s,a_1,a_2),
\end{split}
\end{equation}
where $\hat{\eta'}_{\text{full}}=\eta' C_1(s) + \eta'^2 C_2(s) + O(\eta'^3)$. In order to prove Theorem \ref{th:projection_performance}, we need to find the values of $\hat{\eta}_{\text{1st}}$ that yields:
\begin{equation}
\label{eq:th3_1}
\begin{split}
    \hat{\eta}_{\text{1st}}&=\hat{\eta'}_{\text{full}}\\
    &=\eta' C_1(s) + \eta'^2 C_2(s) + O(\eta'^3),
\end{split}
\end{equation}
and at the same time $\eta'\in \mathbb{R}^+$.
By neglecting $O(\eta'^3)$ and given that $\hat{\eta}_{\text{1st}} \in \mathbb{R}^+$, there are two cases to be considered:
\begin{itemize}
    \item if $C_2(s)$ is non-negative, then for any value of $\hat{\eta}_{\text{1st}} \in \mathbb{R}^+$, there exists $\eta \in \mathbb{R}^+$.
    \item if $C_2(s)$ is negative, then for $\hat{\eta}_{\text{1st}} < \frac{C_1(s)^2}{4|C_2(s)|}$, there exists $\eta \in \mathbb{R}^+$.
\end{itemize}
Therefore, for sufficiently small $\hat{\eta}_{\text{1st}}$, i.e., $\hat{\eta}_{\text{1st}} < \frac{C_1(s)^2}{4|C_2(s)|}$, there always exists $\eta \in \mathbb{R}^+$, and consequently, Theorem \ref{th:projection_performance} is proved.

\subsection{Proof of Theorem \ref{th:time}}
\label{apsec:Proof of Theorem time}

As both policy and state-action value functions are approximated via neural networks, the time complexity of the gradient anticipation follows the time complexity of backpropagation in neural networks. As assumed, the policy and state-action value networks have the same number of hidden layers, $H$, and neurons in each hidden layer, $N$. Therefore, the backpropagation time complexity of the networks for an input state of size $N_s$ and action of size $N_a$ is \citep{lister1995empirical}:
\begin{itemize}
    \item Backpropagation time complexity in the policy network: $O(N_s N + (H-1) N^2 + N N_a)$
    \item Backpropagation time complexity in the state-action value network: $O((N_s+N_a)N + (H-1) N^2 + N)$
\end{itemize}
Given that $N > N_s+N_a$, the time complexity of both networks can be upper bounded by $O(L N^2)$ where we defined $L=H+1$. Now we assume that agent $i\in\mathcal{N}$ wants to anticipate the learning step of another agent $j\in\{\mathcal{N}-\{i\}\}$. In the case of policy parameter anticipation, agent $i$ anticipates $\Delta \theta_j(s)$ as:
\begin{equation}
\begin{split}
    \Delta\theta_j(s) &=\eta \nabla_{\theta_j}\mu_{\theta_j}(s)\nabla_{a_j}Q_j(s,a_1,...,a_n)|_{a_j=\mu_{\theta_j}(s)}.
\end{split}
\end{equation}
Therefore, the time complexity is $O(L N^2)\times O(L N^2)$, or in other words, $O(L^2 N^4)$. In the case of action anticipation, on the other hand, agent $i$ anticipates $\Delta a_j(s)$ as:
\begin{equation}
\begin{split}
    \Delta a_j &= \hat{\eta}_{\text{1}st}\nabla_{a_j}Q_2(s,a_1,...,a_n),
\end{split}
\end{equation}
which has the complexity of $O(L N^2)$. Consequently, the time complexity is reduced by $O(L N^2)$, and Theorem \ref{th:time} is proved.

\begin{algorithm}[H]
\begin{algorithmic}
    \STATE Initialize $\mu_{\theta_i}$, $Q_i$, $\mu'_{i}$, and $Q'_i$    $\forall i \in \mathcal{N}$, and set $\hat{\eta}_{\text{1st}}$
    \FOR{$\textrm{episode}=1\textrm{ to max-num-episodes}$}
        \STATE Receive initial state $s$
        \FOR{$t=1\textrm{ to max-episode-length}$}
            \STATE Select action $a_i$ from $\pi^b_{\theta_i}(s)$ $\forall i \in \mathcal{N}$
            \STATE Execute actions $a=\{a_i\}_{\forall i \in \mathcal{N}}$ and observe rewards $r=\{r_i\}_{\forall i \in \mathcal{N}}$ and new state $s'$
            \STATE Store the tuple $(s,a,r,s')$ in replay buffer $\mathcal{D}$
            \STATE Set $s= s'$
            \STATE
            \STATE Sample a random $K$ tuples $\{(s^k,a^k,r^k,s'^k)\}_{k\in\{1,...,K\}}$ from $\mathcal{D}$
            \FOR{agent $i=1\textrm{ to }n$}
                \STATE Set $y^k_i=r^k_i+\gamma Q'_i(s'^k,a'_1,...,a'_n)|_{a'_h=\mu'_{h}(s'^k)}$, for $k\in\{1,...,K\}$
                \STATE Update state-action value function $Q_i$ by minimizing: $$\mathcal{L}(\omega_i)=\frac{1}{K}\sum_{k\in\{1,...,K\}}[(Q_i(s^k,a^k_1,...,a^k_n;\omega_i)-y_i^k)^2]$$
            \ENDFOR 
            \STATE Set $a^k_i=\mu_{\theta_i}(s^k)$, for $k\in\{1,...,K\}$ and $i \in \mathcal{N}$
            \FOR{agent $i=1\textrm{ to }n$}
                \FOR{agent $j=1\textrm{ to }n$} 
                    \STATE \textbf{if} $j=i$  \textbf{then} continue
                    \STATE Set $\Delta a^k_j=\hat{\eta}_{\text{1st}}\frac{\partial}{\partial a^k_j}Q_j(s^k,a^k_1,...,a^k_n)$ for $k\in\{1,...,K\}$
                \ENDFOR 
            \STATE Update policy parameters $\theta_i$ via:
            $\nabla_{\theta_i}J^{\text{LOLA-OffPA2}}_i =$
            $$ \frac{1}{K}\sum_{k\in\{1,...,K\}} \nabla_{\theta_i}\mu_{\theta_i}(s^k)\frac{\partial}{\partial a^k_i}Q_i(s^k,a^k_1+\Delta a^k_1,...,a^k_i,...,a^k_n+\Delta a^k_{n}) $$
            \ENDFOR  
            \STATE Update $Q'_i$ and $\mu'_i$ $\forall i \in \mathcal{N}$    
        \ENDFOR
    \ENDFOR
\end{algorithmic}
 \caption{LOLA-OffPA2 for a set of $n$ self-interested agents ($\mathcal{N}$).   }
 \label{alg:LOLA-OffPA2}
\end{algorithm}

\begin{algorithm}[H]
\begin{algorithmic}
    \STATE Initialize $\mu_{\theta_i}$, $Q_i$, $\mu'_{i}$, and $Q'_i$    $\forall i \in \mathcal{N}$, and set $\hat{\eta}_{\text{1st}}$
    \FOR{$\textrm{episode}=1\textrm{ to max-num-episodes}$}
        \STATE Receive initial state $s$
        \FOR{$t=1\textrm{ to max-episode-length}$}
            \STATE Select action $a_i$ from $\pi^b_{\theta_i}(s)$ $\forall i \in \mathcal{N}$
            \STATE Execute actions $a=\{a_i\}_{\forall i \in \mathcal{N}}$ and observe rewards $r=\{r_i\}_{\forall i \in \mathcal{N}}$ and new state $s'$
            \STATE Store the tuple $(s,a,r,s')$ in replay buffer $\mathcal{D}$
            \STATE Set $s= s'$
            \STATE
            \STATE Sample a random $K$ tuples $\{(s^k,a^k,r^k,s'^k)\}_{k\in\{1,...,K\}}$ from $\mathcal{D}$
            \FOR{agent $i=1\textrm{ to }n$}
                \STATE Set $y^k_i=r^k_i+\gamma Q'_i(s'^k,a'_1,...,a'_n)|_{a'_h=\mu'_{h}(s'^k)}$, for $k\in\{1,...,K\}$
                \STATE Update state-action value function $Q_i$ by minimizing: $$\mathcal{L}(\omega_i)=\frac{1}{K}\sum_{k\in\{1,...,K\}}[(Q_i(s^k,a^k_1,...,a^k_n;\omega_i)-y_i^k)^2]$$
            \ENDFOR 
            \STATE Set $a^k_i=\mu_{\theta_i}(s^k)$, for $k\in\{1,...,K\}$ and $i \in \mathcal{N}$
            \FOR{agent $i=1\textrm{ to }n$}
                \FOR{agent $j=1\textrm{ to }n$} 
                    \STATE \textbf{if} $j=i$  \textbf{then} continue
                    \STATE Set $\Delta a^k_j=\hat{\eta}_{\text{1st}}\frac{\partial}{\partial a^k_j}Q_j(s^k,a^k_1,...,a^k_n)$ for $k\in\{1,...,K\}$
                \ENDFOR 
            \STATE Update policy parameters $\theta_i$ via:
            $\nabla_{\theta_i}J^{\text{LA-OffPA2}}_i =$
            $$ \frac{1}{K}\sum_{k\in\{1,...,K\}} \nabla_{\theta_i}\mu_{\theta_i}(s^k)\frac{\partial}{\partial a^k_i}Q_i(s^k,a^k_1+\perp\Delta a^k_1,...,a^k_i,...,a^k_n+\perp\Delta a^k_{n}) $$
            \ENDFOR  
            \STATE Update $Q'_i$ and $\mu'_i$ $\forall i \in \mathcal{N}$    
        \ENDFOR
    \ENDFOR
\end{algorithmic}
 \caption{LA-OffPA2 for a set of $n$ self-interested agents ($\mathcal{N}$).   }
 \label{alg:LA-OffPA2}
\end{algorithm}

\begin{algorithm}[H]
\begin{algorithmic}
    \STATE Initialize $\mu_{\theta_i}$, $Q_i$, $\mu'_{i}$, and $Q'_i$    $\forall i \in \mathcal{M}$, and set $\hat{\eta}_{\text{1st}}$
    \FOR{$\textrm{episode}=1\textrm{ to max-num-episodes}$}
        \STATE Receive initial state $s$
        \FOR{$t=1\textrm{ to max-episode-length}$}
            \STATE Select action $a_i$ from $\mu_{\theta_i}(s)$ and the exploration strategy $\forall i \in \mathcal{M}$
            \STATE Execute actions $a=\{a_i\}_{\forall i \in \mathcal{M}}$ and observe common reward $r$ and new state $s'$
            \STATE Store the tuple $(s,a,r,s')$ in replay buffer $\mathcal{D}$
            \STATE Set $s= s'$
            \STATE 
            \STATE Sample a random $K$ tuples $\{(s^k,a^k,r^k,s'^k)\}_{k\in\{1,...,K\}}$ from $\mathcal{D}$
            \STATE Set $y^k=r^k+\gamma Q'(s'^k,a'_1,...,a'_m)|_{a'_h=\mu'_{h}(s'^k)}$, for $k\in\{1,...,K\}$
            \STATE Update state-action value function $Q$ by minimizing: $$\mathcal{L}=\frac{1}{K}\sum_{k\in\{1,...,K\}}[(Q(s^k,a^k_1,...,a^k_m)-y^k)^2]$$
            \STATE Assign the agents into $m$ hierarchy levels using Eq. \ref{eq:hrmaddpg1}
            \STATE Rename the agent assigned to level $i$ as the agent $i$, $\forall i \in \mathcal{M}$
            \STATE Set $a^k_i=\mu_{\theta_i}(s^k)$, for $\forall k\in\{1,...,K\}$ and $\forall i \in \mathcal{M}$
            \FOR{agent $i=m\textrm{ to }1$}
                \FOR{agent $j=1\textrm{ to }i$}
                \STATE Compute $\Delta a^k_j$, for $k\in\{1,...,K\}$: 
                \STATE \textbf{if} $\;\;\,j=1 \;\&\; i\neq m$ \textbf{then} $\Delta a^k_1=\hat{\eta}_{\text{1st}}\frac{\partial}{\partial a^k_1}Q(s^k,a^k_1,...,a^k_{i},\bar{a}^k_{i+1},...,\bar{a}^k_m)$ 
                \STATE \textbf{elif} $j\neq 1 \;\&\; i=m$ \textbf{then} $\Delta a^k_j=\hat{\eta}_{\text{1st}}\frac{\partial}{\partial a^k_j}Q(s^k,a^k_1+\Delta a^k_1,...,a^k_{j-1}+\Delta a^k_{j-1},a^k_j,...,a^k_m)$ 
                \STATE \textbf{elif} $j=1 \;\&\; i= m$ \textbf{then} $\Delta a^k_1=\hat{\eta}_{\text{1st}}\frac{\partial}{\partial a^k_1}Q(s^k,a^k_1,...,a^k_m)$ 
                \STATE \textbf{else} $\Delta a^k_j=\hat{\eta}_{\text{1st}}\frac{\partial}{\partial a^k_j}Q(s^k,a^k_1+\Delta a^k_1,...,a^k_{j-1}+\Delta a^k_{j-1},a^k_j,\bar{a}^k_{j+1},...\bar{a}^k_m)$ 
                \ENDFOR 
            \STATE Update policy parameters $\theta_i$ via:
            $$ \nabla_{\theta_i}J^{\text{HR}}_i \approx \frac{1}{K}\sum_{k\in\{1,...,K\}} \nabla_{\theta_i}\mu_{\theta_1}(s^k)\Delta a^k_i $$
            \STATE Set $\bar{a}^k_i=\text{detach}(a^k_i+\Delta a^k_i)$, for $k\in\{1,...,K\}$
            \ENDFOR  
            \STATE Update $Q'_i$ and $\mu'_i$ $\forall i \in \mathcal{M}$    
        \ENDFOR
    \ENDFOR
\end{algorithmic}
 \caption{HLA-OffPA2 for a set of $m$ common-interested agents ($\mathcal{M}$).   }
 \label{alg:hrlamaddpg}
\end{algorithm}

\section{Implementations details}
\label{apsec:Implementation details}
In this section, we describe the implementations of methods in detail. In order to have fair comparisons between the methods, we have used policies and value functions with the same neural network architecture in all methods. Algorithms \ref{alg:LOLA-OffPA2}, \ref{alg:LA-OffPA2}, and \ref{alg:hrlamaddpg} illustrates the optimization frameworks for LOLA-OffPA2,  LA-OffPA2, and  HLA-OffPA2, respectively.


\textbf{A note on partial observability}. So far, we have formulated the MARL setup as an MG, where it is assumed that the agents have access to the state space. However, in many games, the agents only receive a private state observation of the current state. In this case, the MARL setup can be formulated as a Partially Observable Markov Game (PO-MG) \citep{littman1994markov}. A PO-MG is a tuple $(\mathcal{N},\mathcal{S},\{\mathcal{A}_i\}_{i \in \mathcal{N}},\{\mathcal{O}_i\}_{i \in \mathcal{N}},\{\mathcal{R}_i\}_{i \in \mathcal{N}},\mathcal{T},\{{\Omega}_i\}_{i \in \mathcal{N}},\rho,\gamma)$, where $\mathcal{O}_i$ is the set of sate observations for agent $i \in \mathcal{N}$. Each agent $i$ chooses its action $a_i \in \mathcal{A}_i$ through the policy $\mu_{\theta_i}:\mathcal{O}_i \rightarrow \mathcal{A}_i$ parameterized by $\theta_i$ conditioning on the given state observation $o_i \in \mathcal{O}_i$. After the transition to a new state, each agent $i$ receives a private state observation through its observation function $\Omega_i: \mathcal{S}\rightarrow \mathcal{O}_i$. In this case, the centralized state-action value function for each
agent $i$ is defined as $Q_i(o_1,...,o_n,a_1,...,a_{n})=\mathbb{E}[G_i^t(\tau|s^t=s,o_i=\Omega_i(s)\;\&\;a_i^t=a_i\;\; \forall i \in \mathcal{N})]$. Therefore, the proposed OffPA2 framework can be modified accordingly.

\subsection{Iterated rotational game and iterated prisoner's dilemma}
\label{apsec:Matrix games}
We employed Multi-Layer Perceptron (MLP) networks with two hidden layers of dimension 64 for policies and value functions. In order to make the state-action value functions any-order differentiable, we used SiLU nonlinear function \citep{elfwing2018sigmoid} in between the hidden layers. For IRG, we used the Sigmoid function in the policies to output 1-D continues action, and for IPD, we used the Gumble-softmax function \citep{jang2017categorical} in the policies to output two discrete actions. The algorithms are trained for 900 (in IRG) and 50 (in IPD) episodes by running Adam optimizer \citep{kingma2014adam} with a fixed learning rate of $0.01$. The (projected) prediction lengths in OffPA2 and DiCE frameworks are tuned and set to 0.8 and 0.3, respectively. All experiments are repeated five times, and the results are reported in terms of mean and standard deviation.

\subsection{Exit-Room game}
\label{apsec:Exit-Room game}
Both policy and value networks consist of two parts: encoder and decoder. The encoders are CNN networks with three convolutional layers ($12\times 90\times 90 \rightarrow 32\times 21\times 21 \rightarrow 64\times 9\times 9 \rightarrow 64\times 7\times 7$ ) and two fully connected layers ($3136 \rightarrow 512 \rightarrow 128$ ), with SiLU nonlinear functions \citep{elfwing2018sigmoid} in between. The decoders are MLP networks with two hidden layers of dimension 64 for policies and value functions. We used the Gumble-softmax function in the policies \citep{jang2017categorical} to output the discrete actions. The algorithms are trained for 450 (in level one) and 4500 (in levels two and three) episodes by running Adam optimizer \citep{kingma2014adam} with a fixed learning rate of $0.01$. The (projected) prediction lengths in OffPA2 and DiCE frameworks are tuned and set to 1 and 0.4, respectively. All experiments are repeated five times, and the results are reported in terms of mean and standard deviation.

\subsection{Particle-coordination game}
\label{apsec:Particle coordination game}
We employed MLP networks with two hidden layers of dimension 64 for policies and state-action value functions with SiLU nonlinear functions \citep{elfwing2018sigmoid} in between. We used the Gumble-softmax function \citep{jang2017categorical} in the policies to output the discrete actions. The algorithms are trained for 100k episodes by running Adam optimizer \citep{kingma2014adam} with a fixed learning rate of $0.01$. We set the projected prediction length to $0.1$ for HLA-OffPA2 agents. All experiments are repeated five times, and the results are reported in terms of mean and standard deviation.

\subsection{Standard multi-agent games}
\label{apsec:Standard multi-agent games}
\begin{table}[t]
\begin{minipage}{\linewidth}
  \begin{adjustbox}{width=\linewidth}
    \begin{tabular}{l|  ccc|ccc }
    	\toprule
    	\multicolumn{1}{c}{}   &
    	\multicolumn{3}{c}{$\hat{\eta}_{\text{1st}}$ in Particle Environment}  &
    	\multicolumn{3}{c}{$\hat{\eta}_{\text{1st}}$ in Mujoco Environment}  \\
    	\cmidrule(lr){2-4}
    	\cmidrule(lr){5-7}
    	     & Cooperative Navigation & Physical Deception & Predator-Prey & Half-Cheetah & Walker & Reacher\\
    	\toprule
        HLA-OffPA2 & 0.003    & 0.01 & 0.04 & 0.004 & 0.004 & 0.007 \\
    	\bottomrule
    \end{tabular}
    \end{adjustbox} 
    \captionof{table}{\label{table:prediction_step} The optimized projected prediction lengths for OffPA2-based methods.}
\end{minipage}
\end{table}
As before, we used policies and state-action value functions with the same neural network architecture in all methods. We employed MLP networks with two hidden layers (of dimension 64 for the Particle environment and 256 for the Mujoco environment) for policies and state-action value functions with SiLU nonlinear functions \citep{elfwing2018sigmoid}. In the Particle environment, We used the Gumble-softmax function \citep{jang2017categorical} in the policies to output the discrete actions and trained the algorithms for 100k episodes by running Adam optimizer~\citep{kingma2014adam}~with a fixed learning rate of $0.01$. In the Mujoco environment, we used the Tanh function in the policies to output the continuous actions and train the algorithms for 10k episodes by running Adam optimizer~\citep{kingma2014adam}~with a fixed learning rate of $0.001$. The projected prediction lengths for HLA-OffPA2 agents are optimized between $0.001-0.1$ in all games. The optimized projected prediction lengths are reported in Table \ref{table:prediction_step}.

\vskip 0.2in
\bibliography{sample}

\end{document}